\documentclass[runningheads]{llncs}
\usepackage{graphicx}
\usepackage{algorithm,algpseudocode,float}
\usepackage{lipsum, amsmath, amsfonts}
\usepackage{babel,blindtext}
\usepackage{subcaption}
\algnewcommand\algorithmicforeach{\textbf{foreach}}
\algdef{S}[FOR]{ForEach}[1]{\algorithmicforeach\ #1\ \algorithmicdo}
\algnewcommand\algorithmicinput{\textbf{Input:}}
\algnewcommand\algorithmicoutput{\textbf{Output:}}
\algnewcommand\Input{\item[\algorithmicinput]}%
\algnewcommand\Output{\item[\algorithmicoutput]}%

\makeatletter
\newenvironment{breakablealgorithm}
  {% \begin{breakablealgorithm}
   \begin{center}
     \refstepcounter{algorithm}% New algorithm
     \hrule height.8pt depth0pt \kern2pt% \@fs@pre for \@fs@ruled
     \renewcommand{\caption}[2][\relax]{% Make a new \caption
       {\raggedright\textbf{\ALG@name~\thealgorithm} ##2\par}%
       \ifx\relax##1\relax % #1 is \relax
         \addcontentsline{loa}{algorithm}{\protect\numberline{\thealgorithm}##2}%
       \else % #1 is not \relax
         \addcontentsline{loa}{algorithm}{\protect\numberline{\thealgorithm}##1}%
       \fi
       \kern2pt\hrule\kern2pt
     }
  }{% \end{breakablealgorithm}
     \kern2pt\hrule\relax% \@fs@post for \@fs@ruled
   \end{center}
  }
\makeatother
\usepackage{hyperref}
\usepackage{cleveref}

\begin{document}
\title{A spectral algorithm for finding maximum cliques in dense random intersection graphs \thanks{Christoforos Raptopoulos was supported by the Hellenic Foundation for Research and Innovation (H.F.R.I.) under the “2nd Call for H.F.R.I. Research Projects to support Post-Doctoral Researchers” (Project Number:704). \\ Paul Spirakis was supported by the NeST initiative of the EEE and CS of the U. of Liverpool and by the EPSRC grant EP/P02002X/1.}}
\titlerunning{A spectral algorithm for maximum cliques in RIGs}
% If the paper title is too long for the running head, you can set
% an abbreviated paper title here
%
\author{Filippos Christodoulou\inst{1}\orcidID{0000-0003-3176-3759
} \and
Sotiris Nikoletseas\inst{2,3}\orcidID{0000-0003-3765-5636} \and
Christoforos Raptopoulos\inst{3}\orcidID{0000-0002-9837-2632} \and Paul Spirakis\inst{4,2}\orcidID{0000-0001-5396-3749}}
\authorrunning{F. Christodoulou, S. Nikoletseas, C. Raptopoulos, P. Spirakis}
% First names are abbreviated in the running head.
% If there are more than two authors, 'et al.' is used.
%
\institute{
Gran Sasso Science Institute, L'Aquila, Italy \\ \email{filippos.christodoulou@gssi.it} 
\and Computer Engineering \& Informatics Department, University of Patras, Greece
\email{\{nikole,raptopox\}@ceid.upatras.gr}\\
\and Computer Technology Institute \& Press “Diophantus” (CTI), Patras, Greece 
\and  Department of Computer Science, University of Liverpool, UK\\
\email{P.Spirakis@liverpool.ac.uk}}
\maketitle              % typeset the header of the contribution
\begin{abstract}
In a random intersection graph $G_{n,m,p}$, each of $n$ vertices selects a random subset of a set of $m$ labels by including each label independently with probability $p$ and edges are drawn between vertices that have at least one label in common. Among other applications, such graphs have been used to model social networks, in which individuals correspond to vertices and various features (e.g. ideas, interests) \linebreak correspond to labels; individuals sharing at least one common feature are connected and this is abstracted by edges in random intersection graphs. In this paper, we consider the problem of finding maximum cliques when the input graph is $G_{n,m,p}$. Current algorithms for this problem are successful with high probability only for relatively sparse instances, leaving the dense case mostly unexplored. We present a spectral algorithm for finding large cliques that processes vertices according to respective values in the second largest eigenvector of the adjacency matrix of induced subgraphs of the input graph corresponding to common neighbors of small cliques. Leveraging on the Single Label Clique Theorem from \cite{NRS12}, we were able to construct random instances, without the need to externally plant a large clique in the input graph. In particular, we used label choices to determine the maximum clique and then concealed label information by just giving the adjacency matrix of $G_{n, m, p}$ as input to the algorithm. Our experimental evaluation showed that our spectral algorithm clearly outperforms existing polynomial time algorithms, both with respect to the failure probability and the approximation guarantee metrics, especially in the dense regime, thus suggesting that spectral properties of random intersection graphs may be also used to construct efficient algorithms for other NP-hard graph theoretical problems as well.  

\keywords{Random Intersection Graphs \and Maximum Cliques \and Heuristics}
\end{abstract}
\section{Introduction}
A \emph{clique} in an undirected graph $G$ is a subset of vertices any two of which are connected by an edge. The problem of finding the maximum clique in an arbitrary graph is fundamental in Theoretical Computer Science and appears in many different settings. As an example, consider a social network where vertices represent people and edges represent mutual acquaintance. Finding a maximum clique in this network corresponds to finding the largest subset of people who all know each other. More generally, the analysis of large networks in order to identify communities, clusters,
and other latent structure has come to the forefront of much research. The Internet, social networks, bibliographic databases, energy distribution networks, and global networks of economies are some of the examples motivating the development of the field.

From a computational complexity point of view, it is well known that  \linebreak determining the size of the largest clique of an arbitrary graph of $n$ vertices is NP-complete \cite{K72}. This fact is further strengthened in \cite{CHKX06}, showing that, if $k$ is the size of the maximum clique, then the clique problem cannot be solved in time $n^{o(k)}$, unless the exponential time hypothesis fails. Additionally, there are several results on hardness of approximation which suggest that there can be no approximation algorithm with an approximation ratio significantly less than linear (see e.g. \cite{H99}).

The intractability of the maximum clique problem for arbitrary graphs lead researchers to the study of the problem for appropriately generated random graphs. In particular, for Erd\H{o}s-R\'enyi random graphs $G_{n, \frac{1}{2}}$ (i.e. random graphs of $n$ vertices, in which each edge appears independently with probability $\frac{1}{2}$), there are several greedy algorithms that find a clique of size about $\log_2{n}$ with high probability (whp, i.e. with probability that tends to 1 as $n$ goes to infinity), see e.g. \cite{GM75,K76}. Since the clique number of $G_{n, \frac{1}{2}}$ is asymptotically equal to $2 \log_2{n}$ with high probability, these algorithms approximate the clique number by a factor of 2. It has been conjectured that finding a clique of size $(1+\Theta(1)) \log_2{n}$, in a random graph instance $G_{n, \frac{1}{2}}$, in which we have planted a randomly chosen clique of size $n^{0.49}$, with at least constant probability, would require techniques beyond the current limits of complexity theory. 
%This belief was strengthened by the fact that the Metropolis algorithm can take super-polynomial time to locate a clique in $G_{n, \frac{1}{2}}$ that is only slightly bigger than that produced by greedy heuristics (see \cite{J92}). A more dramatized version of the above conjecture was presented in \cite{J92}, stating that the problem of finding an $1.01 \log_2{n}$ clique remains hard even if the input graph is a $G_{n, \frac{1}{2}}$ random graph in which we have planted a randomly chosen clique of size $n^{0.49}$. 
This conjecture seems to identify a certain bottleneck for the problem; finding the maximum clique in the case where the planted clique has size at least $\sqrt{n}$ can be done in polynomial time by using spectral properties of the adjacency matrix of the graph (see \cite{AKS98}). %We finally note that there are quite a few nice results concerning generalizations of the planted clique problem in various (quite general) random graphs models (see e.g. \cite{C06, CL09}).    

In this paper, we consider random instances of the random intersection graphs model (introduced in \cite{KSS99,S95}) as input graphs. In this model, denoted by ${\cal G}_{n, m, p}$, each one of $m$ labels is chosen independently with probability $p$ by each one of $n$ vertices, and there are edges between any vertices with overlaps in the labels chosen. One of the most interesting results regarding this model is that, when the number of labels is sufficiently large (in particular, when $m=n^{\alpha}, \alpha \geq 3$) the random intersection graphs model is equivalent to the Erd\H{o}s-R\'enyi random graphs model (in the sense that the total variation distance between the two spaces tends to 0; see \cite{FSS00,R11}). Random intersection graphs are relevant to and capture quite nicely social networking. Indeed, a social network is a structure made of nodes (individuals or organizations) tied by one or more specific types of interdependency, such as values, visions, financial exchange, friends, conflicts, web links etc. Social network analysis views social relationships in terms of nodes and ties. Nodes are the individual actors within the networks and ties are the relationships between the actors. Other applications include oblivious resource sharing in a (general) distributed setting, efficient and secure communication in sensor networks \cite{NRS11}, interactions of mobile agents traversing the web etc. For recent research related to random intersection graphs we refer the interested reader to the surveys \cite{survey1,survey2} and references therein.

\subsection{Previous work on maximum cliques in random intersection graphs.}

In \cite{S95}, the authors used the first moment probabilistic method to provide a lower bound on the clique number of random instance of ${\cal G}_{n, m, p}$ in the case where $mp^2$ tends to a constant as $n \to \infty$. In \cite{NRS12,TCS21} this range of values was considerably extended and a precise characterization of maximum cliques was given in the case where $m = n^{\alpha}, \alpha<1$ and $p=O(m^{-1/2})$. In particular, the Single Label Clique Theorem was proved, indicating that, with high probability any clique $Q$ of size $|Q| \sim np$ in a random instance of ${\cal G}_{n, m, p}$ (and thus also the maximum clique) is formed by a single label. 
However, these structural results are existential and thus do not lead to algorithms for finding the maximum clique. It is worth noting that, the equivalence results between the random intersection graphs model and the Erd\H{o}s-R\'enyi random graphs model for large number of vertices suggest that the problem of finding a maximum clique in a random instance of ${\cal G}_{n, m, p}$ in this range of values should not be any easier in the former that it is in the latter. On the other hand, in the range of values $m=n^{\alpha}, \alpha<1$ and $p=O(m^{-2/3})$, greedy algorithms for finding large cliques in random intersection graphs were presented in the work \cite{BK17}. The first algorithm in that paper, referred as GREEDY-CLIQUE, finds a clique of the optimal order in a random instance of ${\cal G}_{n, m, p}$ with high probability, in the case where the asymptotic degree distribution is a power-law with exponent within $(1,2)$. The algorithm considers vertices in decreasing order of degree and greedily constructs a clique by extending by a vertex if and only if the latter is connected to all other vertices already included; it can be implemented to run in expected time $O(n^2)$. In the same paper \cite{BK17}, in the case where the input graph is a random instance of ${\cal G}_{n, m, p}$ with bounded degree variance, a second greedy algorithm, named MONO-CLIQUE, was suggested, which can be implemented to run in expected time $O(n)$. The main idea of this algorithm is to try and construct a large clique directly by considering common neighbours of endpoints of every edge in the graph. The pseudocodes for GREEDY-CLIQUE and MONO-CLIQUE can be found in Appendixes \ref{sec:greedy-clique} and \ref{sec:mono-clique} respectively.

In \cite{BT06} a more general greedy algorithm was presented, namely the Maximum-Clique Algorithm, which constructs a large clique by considering the common neighborhood of vertex subsets of fixed size $k$ (i.e. independent of $n$) and checking whether it forms a clique. From the cliques found in this way, it takes the largest ones in order to cover the graph. This algorithm finds maximum cliques whp for a wider range of parameters of the model (but still within the sparse regime) than both algorithms GREEDY-CLIQUE and MONO-CLIQUE, at the cost of larger running time. In particular, the Maximum-Clique Algorithm outputs a maximum clique in a random instance of ${\cal G}_{n,m,p}$ with $m=n^\alpha, \alpha < 1$ and $\ln^2 n /n \leq p=O(m^{-2/3})$, with high probability. Since in this paper we consider metrics regarding the ability of an algorithm to find large cliques (namely failure probability and approximation guarantee), we use the Maximum-Clique Algorithm as a benchmark in relation to which we evaluate our spectral algorithm. In fact, we use a slightly more efficient version where we directly exclude $k$-subsets of vertices that are not complete, in order to significantly reduce the $\binom{n}{k}$ factor corresponding to the number of all $k$-sets in the running time. The pseudocode of the benchmark algorithm is shown in Appendix \ref{sec:maximum-clique}. Different pruning ideas for reducing the running time of greedy algorithms for finding large cliques through the reduction of the size of the input graph, have been considered in \cite{FH15}.

\section{Our contribution}
%\vspace{-0.5cm}
In this paper we consider the problem of finding maximum cliques when the input graph is $G_{n,m,p}$. We present a spectral algorithm for finding large cliques that processes vertices according to respective values in the second largest eigenvector of the adjacency matrix of carefully selected induced subgraphs of the input graph created by common neighborhoods of small (constant size) $k$-cliques. Because of the computation of the spectral decomposition, the running time of our algorithm is larger than greedy algorithms in the relevant literature, but it succeeds with higher probability in finding large cliques. In particular, we compared our algorithm with the most efficient version of the Maximum-Clique Algorithm from \cite{BT06}. Leveraging on the Single Label Clique Theorem from \cite{NRS12}, we were able to avoid the construction of artificial input graph instances with known planted large cliques. In particular, we used label choices to determine the maximum clique and then concealed label information by just giving the adjacency matrix of $G_{n, m, p}$ as input to the algorithm. Our experimental evaluation showed that, as we move from sparse instances to denser ones, both metrics regarding the failure probability of our algorithm as well as the approximation guarantee (when the maximum clique is not found) are much better than the corresponding values for the Maximum-Clique algorithm. This difference is \linebreak especially highlighted as we move from sparser instances to denser ones, in which there is no guarantee that greedy algorithms will succeed (but the Single Label Clique Theorem still holds) and also as the size of the $k$-cliques used for creating induced subgraphs increases (yet remains a relatively small constant, e.g. $k=6,7,8$). We believe that our current paper suggests that spectral properties of random intersection graphs may be used to construct efficient algorithms for other NP-hard graph theoretical problems as well.

\section{Definitions, notation and useful results}

Given an undirected graph $G$, we denote by $V(G)$ and $E(G)$ the set of vertices and the set of edges respectively. Edges of $G$ will be denoted as 2-sets; two vertices $v,u$ are connected in $G$ if and only if $\{u,v\} \in E(G)$. For any vertex $v \in V(G)$, we denote by $N(v)=N_G(v)$ the set of neighbours of $v$ in $G$, namely $N(v)\stackrel{\text{def}}{=} \{u \in V(G): \{u,v\} \in E(G)\}$. In addition, for any subset of vertices $S \subseteq V$, we denote by $N(S)$ the set of vertices having at least one neighbor in $S$. We denote by $\deg(v) =|N(v)|$ the degree of $v$. For any subset $S \subseteq V$, we denote by $G[S]$ the induced subgraph of $G$ on $S$, namely $G[S] = (S, \{\{u, v\} \in E(G):v, u \in S\})$. 
Given an arbitrary ordering of the vertices, say $v_1, v_2, \ldots, v_{|V|}$, the adjacency matrix $A_G$ of $G$ is an $|V|\times|V|$ matrix where $A_G[i,j]=1$ if $\{v_i,v_j\} \in E(G)$ and $A_G[i,j]=0$ otherwise. An eigenvector of $\mathbf{A}_G$ with corresponding eigenvalue $\lambda$ is a vector $\mathbf{x}$ for which $\mathbf{A}_G \mathbf{x} = \lambda \mathbf{x}$. Since by definition $\mathbf{A}_G$ is symmetric, it has $|V|$ real eigenvalues $\lambda_1 \geq \lambda_2 \geq \cdots \geq \lambda_{|V|}$, with orthogonal corresponding eigenvectors $\mathbf{x}^{(1)}, \mathbf{x}^{(2)}, \ldots, \mathbf{x}^{(|V|)}$.

%We are going to use the following notation, for the set $A \cup \{x\}$ we write $A+x$. Denote by $\Gamma(v)$ the set of vertices having edges to $v$ and by $N(v) := \Gamma(v) + v$ the same set including v itself. For a vertex $U$ we denote $Z(U)$ the common neighborhood of the vertices in $U(Z(U) := \cap_{k=1}^{i}  N(v_i))$.

The formal definition of the random intersection graphs model is as follows:

\begin{definition}[Random Intersection Graph - ${\cal G}_{n, m, p}$ \cite{KSS99,S95}]
Consider a universe ${\cal M} =\{1, 2, \ldots, m\}$ of labels and a set of $n$ vertices $V$. Assign \linebreak independently to each vertex $v \in V$ a subset $S_{v}$ of ${\cal M}$, choosing each element $\ell \in {\cal M}$ independently with probability $p$ and draw an edge between two vertices $v \neq u$ if and only if $S_{v} \cap S_{u} \neq \emptyset$. The resulting graph is an instance $G_{n, m, p}$ of the random intersection graphs model. 
\end{definition}

In this model we also denote by $L_{\ell}$ the set of vertices that have chosen label $\ell \in M$. Given $G_{n, m, p}$, we refer to $\{L_{\ell}, \ell \in {\cal M}\}$ as its \emph{label representation}. Furthermore, the bipartite graph with vertex set $V \cup {\cal M}$ and edge set $\{(v, \ell): \ell \in S_{v}\} = \{(v, \ell): v \in L_{\ell}\}$ is the \emph{bipartite random graph $B_{n, m, p}$ associated to $G_{n, m, p}$}. Notice that the associated bipartite graph is uniquely defined by the label representation. 

Given a graph $G$, a \emph{clique} is a set of vertices every two of which are connected by an edge; the size of the maximum clique in $G$ is its \emph{clique number}. Notice that, by definition, for any $\ell$ the set of vertices within $L_{\ell}$ forms a clique in $G_{n, m, p}$. Furthermore, the expected size of such a clique is $\mathbb{E}[|L_{\ell}|]=np$. Observe that edges of cliques in $G_{n, m, p}$ may be formed by different labels. However, when the number of labels is smaller than the number of vertices, the following theorem states that, under mild conditions, with high probability, in any large enough clique of $G_{n, m, p}$, edges are formed by a single label.

\begin{theorem}[Single Label Clique Theorem \cite{TCS21}] \label{theorem-slct}
Let $G_{n, m, p}$ be a random instance of the random intersection graphs model with $m = n^{\alpha}, 0<\alpha<1$ and $mp^2 = O(1)$. Then whp, any clique $Q$ of size $|Q| \sim np$ in $G_{n, m, p}$ is formed by a single label. In particular, the maximum clique is formed by a single label.
\end{theorem}

Leveraging on the above theorem, in our experiments we avoid the artificial construction of graph instances with planted cliques. In particular, during the construction of the random intersection graph, we use its label representation to find a set $L_{\ell}$ of maximum cardinality; by the above theorem, this will correspond to a maximum clique, and its size will be the clique number of $G_{n, m, p}$ with high probability. Subsequently, we hide the label representation and give the constructed graph $G_{n, m, p}$ as input to the algorithms that we consider in our experimental evaluation (i.e. just the vertex and edge sets).

\subsection{Range of values for $m,n,p$} 

It follows from the definition of the model that the edges in $G_{n, m, p}$ are not independent. In particular, the (unconditioned) probability that a specific edge exists is $1-(1-p^2)^m$. Therefore, when $mp^2=o(1)$, the expected number of edges of $G_{n, m, p}$ is $(1+o(1)) {\binom{n}{2}} mp^2$. For the range of values $p=O(m^{-2/3}), m=n^{\alpha}, \alpha<1$, where the Maximum-Clique Algorithm of \cite{BT06} is guaranteed to output a large enough clique whp, this becomes $O(n^2 m^{-1/3}) = O(n^{2-\alpha/3})$. On the other end, when $mp^2=\omega(1)$ then the graph is almost complete. In view of this, we will refer to the range of values $m=n^{\alpha}, \alpha<1, mp^2=\Omega(m^{-2/3})$ as the dense regime, noting that the Single Label Clique Theorem continues to hold in this range of values.

\section{The spectral algorithm}

We can now give the details of \emph{Spectral-Max-Clique} algorithm. Inspired by the algorithm in \cite{AKS98}, our algorithm takes as input the graph $G_{n,m,p}$, the size $k$ of a witness $k$-clique (i.e. a small clique that is assumed to belong to the maximum clique) and a parameter $t$, which is a lower bound on the maximum clique size in $G_{n, m, p}$ (recall that, by Theorem \ref{theorem-slct}, when $m=n^{\alpha}, \alpha<1$, any clique $Q$ with size $\lvert Q \rvert \sim np$ in $G_{n,m,p}$ is formed only by a single label; since $np$ is the expected size of $\mathbb{E}[|L_{\ell}|]$, for any $\ell \in {\cal M}$, we set $t = np$). The main difference between our algorithm and the algorithm of \cite{AKS98} is a kind of preprocessing on the input graph, which is done at step 3 of the algorithm; in particular, since the input graph $G_{n, m, p}$ has many large cliques of size $np$ (in fact, by Theorem \ref{theorem-slct}, it has exactly $m$ whp), we work on the induced graph $H$ which has fewer (ideally exactly one) large cliques, namely the ones including $S$. 

At the beginning of the execution, we initialize an empty set $M$, which at the end of the execution will be the output of the algorithm (the maximum clique of the graph). The algorithm enters a for-loop to be repeated as many times as the number of subsets $S \subseteq V$ of size $k$ in the $G_{n,m,p}$. At every iteration of the for-loop, we construct the induced graph $H$ which contains all the vertices of the subset $S$ of the original graph $G_{n,m,p}$ as well as all the neighbors of the vertices in subset $S$ (namely $H = G[S \cup N(S)]$). We then find the adjacency matrix $\mathbf{A}_H$ of $H$ and we find the eigenvector corresponding to the second largest eigenvalue, namely $\mathbf{x}^{(2)}$; the latter can be done in polynomial time. The algorithm then sorts the vertices of $H$ in decreasing order of the absolute values of the corresponding coordinates in the second eigenvector $\mathbf{x}^{(2)}$, where equalities are broken arbitrarily. Subsequently, we consider only the first $t$ vertices in this ordering and store them in an empty set $W$. We then define an empty set $Q$, where the clique (not necessarily the maximum) will be added. Afterward, for every vertex $v$ in $H$, the algorithm checks whether $v$ has at least $3np/4$ neighbors in the set $W$ and exactly $\lvert Q \rvert$ neighbors in the set $Q$. If the two conditions are true, $v$ is included in the set $Q$. In the end, we check if the size of the newly added clique $Q$ is largest from the size of the existing clique in set $M$ ($\lvert Q \rvert \ge \lvert M \rvert$) and finally the maximum clique $M$ is returned by the algorithm. The main heuristic idea why this algorithm works as intended is that, most of the time, the second eigenvector $\mathbf{x}^{(2)}$ of $\mathbf{A}_H$ can be used to find a big portion of the largest clique; intuitively this happens because the maximum clique will be by far the largest most dense induced subgraph of the graph, and this will be depicted in the (absolute) values of the corresponding positions of the second eigenvector (in the extreme case where the $n$-vertex graph consists only of a $k$-sized clique $Q$, the only positions where an eigenvector corresponding to the second largest eigenvalue can have non-zero elements is on the positions corresponding to the vertices of $Q$); see also \cite{AKS98} for a theoretical explanation why this heuristic works in the planted clique model.   Therefore, since the algorithm checks all of the subsets of $V$ of size $k$, in some step it will reach a subset $S$, which belongs to the maximum clique $M$. Our experimental evaluation shows that at this iteration our algorithm succeeds in finding the largest clique of the graph in most cases. The pseudocode of our algorithm is shown below.

%The following algorithm finds cliques of size at least $np$ in a random intersection graph, when $\alpha < 1$. From this clique the algorithm outputs the largest clique. We can see that with high probability of  a graph $G_{n,m,p}$, $m = n^\alpha, 0 < \alpha < 1, p = \Omega(\sqrt{\frac{1}{nm}})$ and $mp^2=O(1)$, the largest clique $Q$ of the graph has size $|Q| \sim np$.

\begin{breakablealgorithm}
   \caption{Spectral-Max-Clique}
   \label{alg-1}
  \begin{algorithmic}[1]
    \Input Random instance of ${\cal G}_{n,m,p}$, parameters $k \in \mathbb{N}$, $t = np$\
    \Output Clique $M$ of $G_{n,m,p}$\
    \State $M=\emptyset$;\
    \ForEach{subset $S \subseteq V, |S| = k$}
        \State Construct the induced graph $H = G[S \cup N(S)]$;\ 
        \State Compute the eigenvector $\mathbf{x}^{(2)}$ corresponding to the second largest \linebreak eigenvalue of $\mathbf{A}_H$;\
        \State Sort the vertices of $H$ in decreasing order of the absolute           values of corresponding coordinates in $\mathbf{x}^{(2)}$;\
        \State Let $W$ be the first $t$ vertices in this ordering;\
        \State Set $Q=\emptyset$;\
        \ForEach{$v \in H$} \qquad %\%Assume arbitrary ordering of vertices
            \If{$v$ has at least $3np/4$ neighbors in $W$ and $|Q|$ neighbors in $Q$}
                \State $Q=Q \cup \{v\}$;\
            \EndIf
        \EndFor
        \If{$|Q|>|M|$}
            \State Set $M=Q$;\
        \EndIf
    \EndFor
    \State \Return $M$
\end{algorithmic}
\end{breakablealgorithm}

\subsection{Running time of our algorithm}

%Time analysis + explanation why we remove the $n^k$ factor in the experiments.

We note that the outer for-loop of our algorithm runs for $\binom{n}{k}$ times. Furthermore, for a given $k$-set $S$, steps 3, 4 and 5 take $O(n^3)$ time, with step 4 regarding spectral decomposition being the most expensive (in theory, spectral decomposition can be done more efficiently in $O(n^{2.4})$, but here we use the time complexity of most practical implementations). Finally, it is easy to see that the for-loop in steps 8 to 12 runs in $O(n^3)$ time, while all other steps are either direct assignments, definition of easily checked conditions and thus take $O(1)$ time. Overall, the running time of our algorithm is $\binom{n}{k} \cdot O(n^3)$. Clearly the most time consuming factor is the number of times that the outer for-loop is running in order to find a good enough starting $k$-set $S$. A similar situation arises also in the algorithm Maximum clique (see Appendix \ref{sec:maximum-clique}), whose running time is $\binom{n}{k} \cdot O(n^2)$. To allow for the algorithms considered and evaluated in our paper to run for larger values of $k$ in the experimental evaluation, we assume, without loss of generality, that a suitable $k$-set is known from the start, thus avoiding the $\binom{n}{k}$ factor in the running time. This is where Theorem \ref{theorem-slct} becomes useful, since a suitable $k$-set $S$ can be any subset of a (single label) maximum clique.

%\vspace{-0.05cm}
\section{Experimental evaluation}

This section is devoted to the presentation of our experimental results regarding the comparison of the algorithms \emph{Spectral-Max-Clique} and \emph{Maximum-Clique} with respect to two metrics, namely failure probability and approximation \linebreak guarantee. In particular, the failure probability is defined as the probability that an algorithm fails to find the maximum clique; in our experimental evaluation this probability is approximated by the fraction of the number of independent instances of random intersection graphs where an algorithm fails to find a \linebreak maximum clique. The approximation guarantee is defined as the fraction of the clique found by an algorithm over the size of a maximum clique; in our experimental evaluation this is approximated by the average of the corresponding fractions achieved by an algorithm for various independent instances of random intersection graphs.

The number of $G_{n,m,p}$ graph instances that have been given as input to the algorithms for small values of $k$ $(k=1,2,3)$ were 2000. However, it is worth noting that, as we increase the value of $k$, the computational resources required also increases, because the dependency on $k$ becomes more prevalent even for straightforward steps of the algorithm \ref{alg-1} (as for example in step 3 for constructing the induced subgraph $H$). In particular, the number of independent graph instances used for $k=4,5,6,7,8$ were 1600, 1400, 800, 700, 500  respectively.

This section is devoted firstly to show our experimental results between the two algorithms for finding the maximum clique that we have considered and secondly we show an approximation guarantee (denoted by \textit{fraction} variable in the \cref{frac1,frac2,frac3,frac4,frac5,frac6}) of the output maximum clique found by each algorithm over the maximum clique of the graph. We present a comparison between \emph{Spectral-Max-Clique} and \emph{Maximum-Clique} algorithm. Further \linebreak experimental results are also shown in the Appendix \ref{sec:further-experiments}.
%Further experimental results can be found in the full version of our paper \cite{Arxiveref}.

Our computing platform is a machine with AMD Ryzen Threadripper 3970X at 3.7GHz, 32 cores, 256 GB RAM, GPU with 2x NVIDIA GeForce RTX 3080 10GB and running Ubuntu Linux version 20.04.2 LTS. The code has been written in Python 3. The code for repeating the experiments is available \href{https://github.com/filipposchr/SpectralMaxClique}{here}.

The goal of our experimental evaluation is to verify whether the \emph{Spectral-Max-Clique} algorithm performs better, meaning that it succeeds to find the maximum clique in a $G_{n,m,p}$, in comparison with the \emph{Maximum-Clique}. As we already mentioned, this happens for dense graphs. In our experiments, we ran both algorithms for different instances of $G_{n,m,p}$. Specifically, we use three different values of parameter $\alpha$, with $\alpha = 1/3, 2/3, 1$ and for the number of nodes ($n$), we set $n$ equal to $1000$ and $3000$. Regarding parameter $p$, we cover a different range of values for each experiment in order to test sparse instances as well as dense ones. It is important to explain how we understand that the output of each algorithm is actually the maximum clique in the graph. We take a $G_{n, m, p}$ instance with the above parameters and find the heaviest label (i.e. the label with the largest number of vertices), call it $\ell_{\text{max}}$. By Theorem \ref{theorem-slct}, the set of vertices in $L_{\ell_{\text{max}}}$ form a maximum clique in $G_{n,m,p}$ whp, so in this way there is no need to externally plant a known large enough clique in the graph. Then, we run the algorithms on this instance, but we give only the graph as input (i.e. the algorithm is unaware of the specific label choices).  We say that the algorithms fail if they do not find a clique at least as large as $|L_{\ell_{\text{max}}}|$. Note that this is a strict condition, namely, even finding a clique of size $|L_{\ell_{\text{max}}}|-1$ is considered a failure. In each case, we gradually increase the selection probability $p$, in order to highlight that the failure probability curve of \emph{Spectral-Max-Clique} is much lower than the failure probability curve of \emph{Maximum-Clique}, especially when the input graphs become denser.

It is worth noting that, the selection of a correct starting set $S$ of $k$ vertices in Step 1 of \emph{Spectral-Max-Clique} pseudocode, implies a multiplicative $\Theta(n^k)$ factor on the running time of our algorithm. Even though for constant $k$ this remains polynomially bounded, in order to allow our experiments to run for large values of $n$ and $k$, we have  assumed that the initial set of $k$ vertices is always chosen from those in the maximum clique.

%\vspace{-0.5cm}
\begin{figure*}[ht]
    \centering
    \begin{subfigure}[b]{0.482\textwidth}
        \centering
        \includegraphics[width=\textwidth]{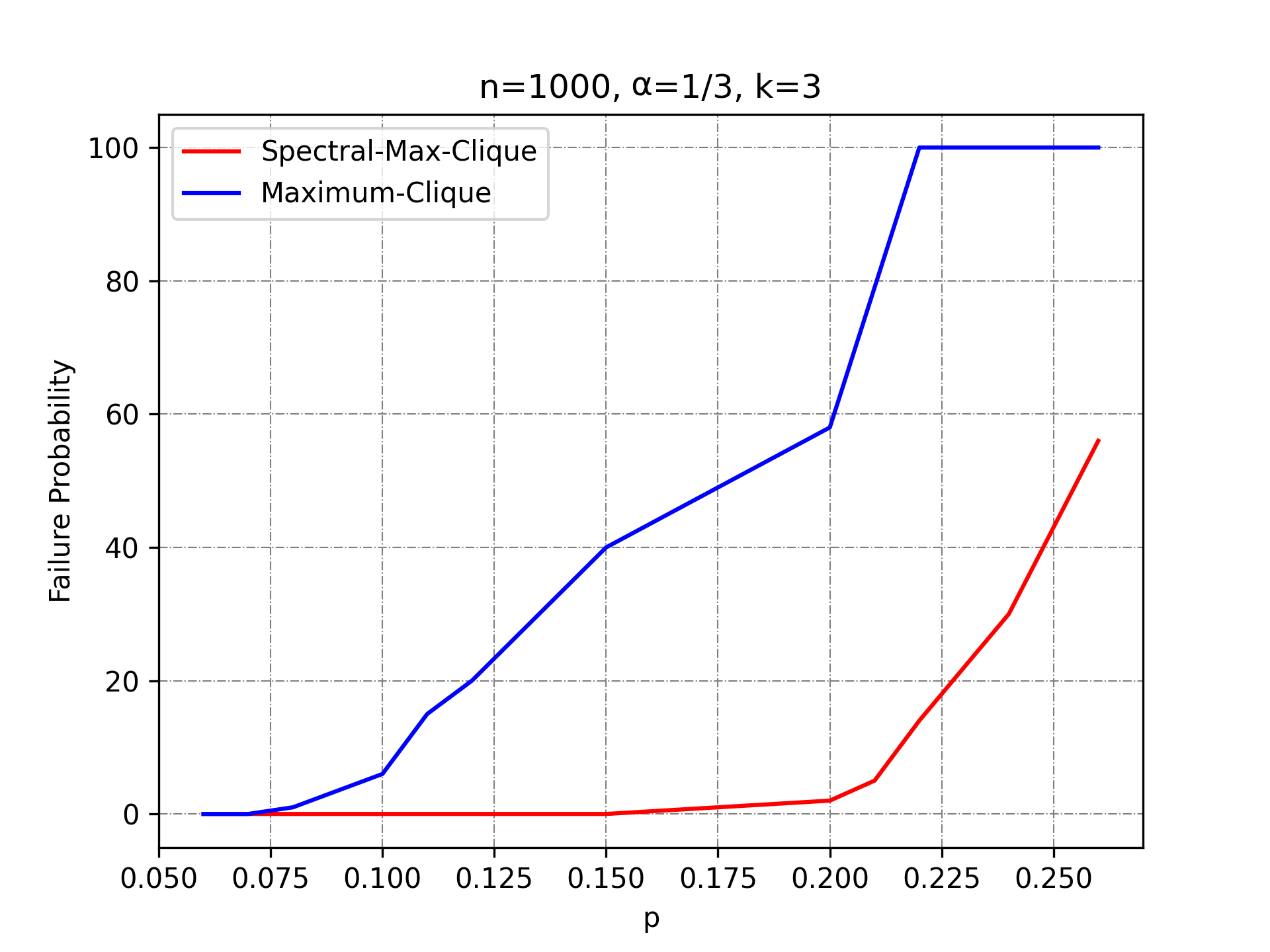}
        \caption{$k = 3$}\label{fig1a}
    \end{subfigure}
    \quad
    \begin{subfigure}[b]{0.482\textwidth}  
        \centering 
        \includegraphics[width=\textwidth]{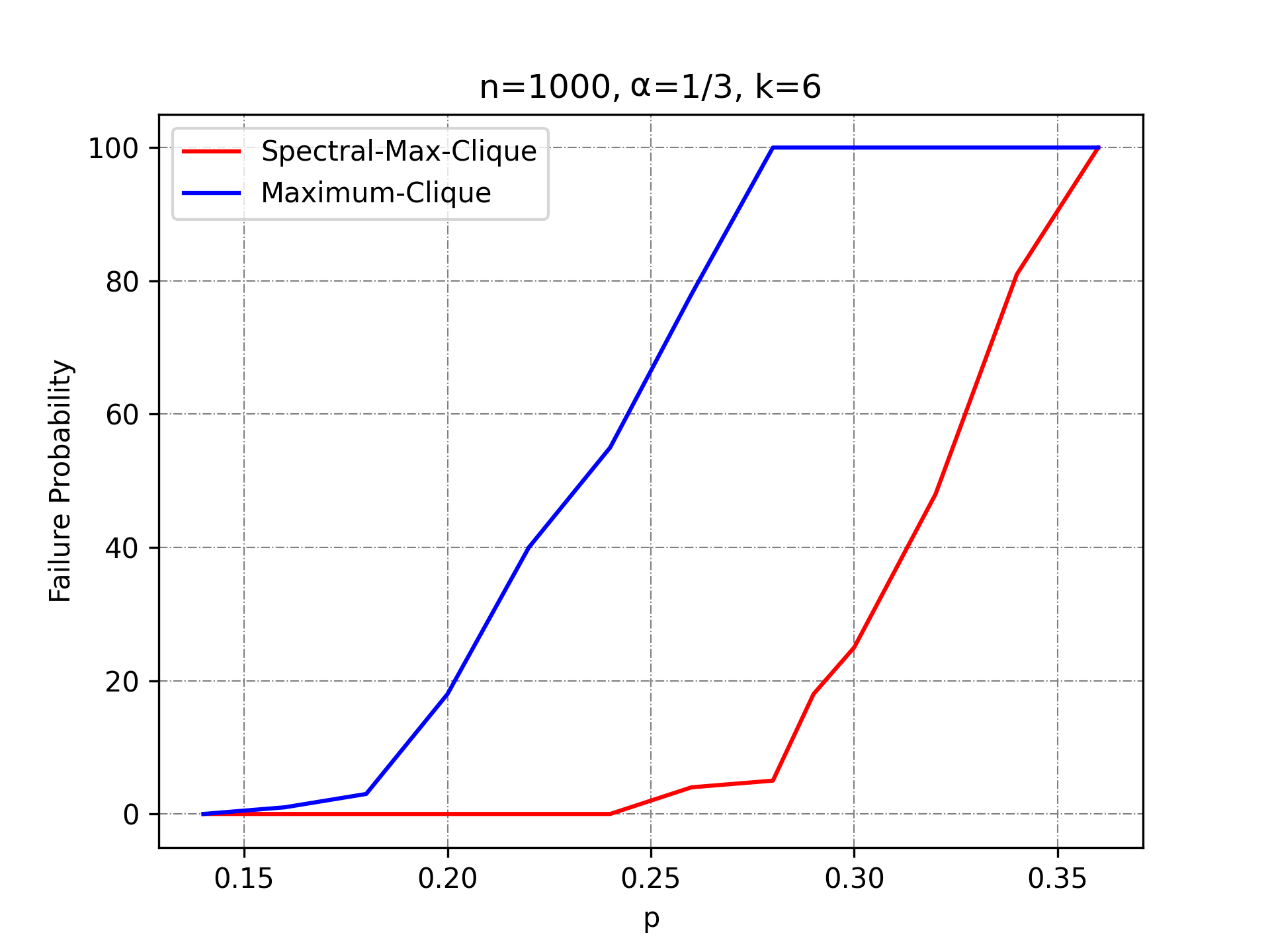}
        \caption{$k = 6$}\label{fig1b}
    \end{subfigure}
    
    \caption{Failure probability curves for $\alpha = 1/3$, $n = 1000$ and $k=3, 6$.}\label{fig1}
\end{figure*}

\begin{figure*}[ht]
    \centering
    \begin{subfigure}[b]{0.482\textwidth}
        \centering
        \includegraphics[width=\textwidth]{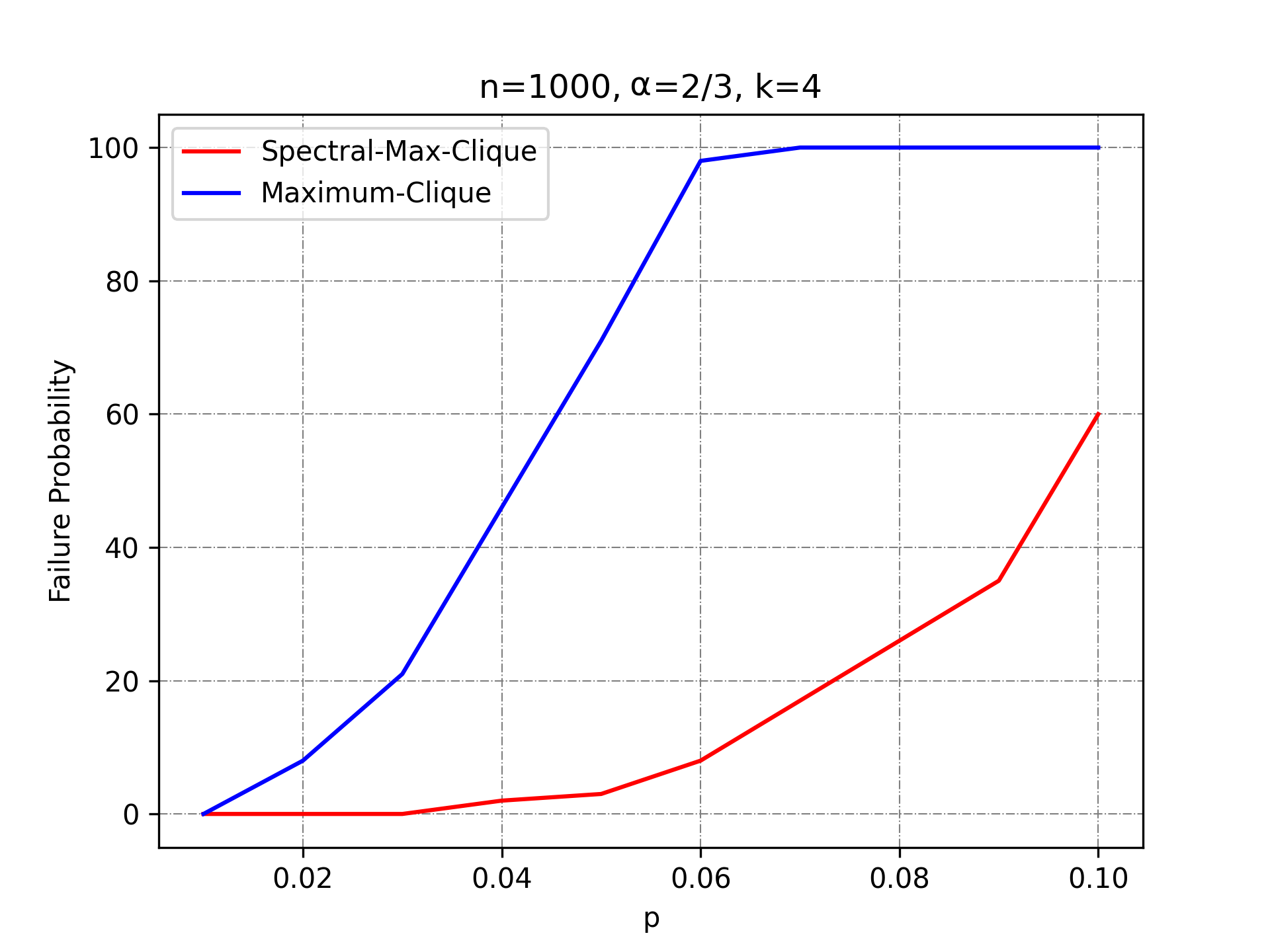}
        \caption{$k = 4$}\label{fig2a}
    \end{subfigure}
    \quad
    \begin{subfigure}[b]{0.482\textwidth}  
        \centering 
        \includegraphics[width=\textwidth]{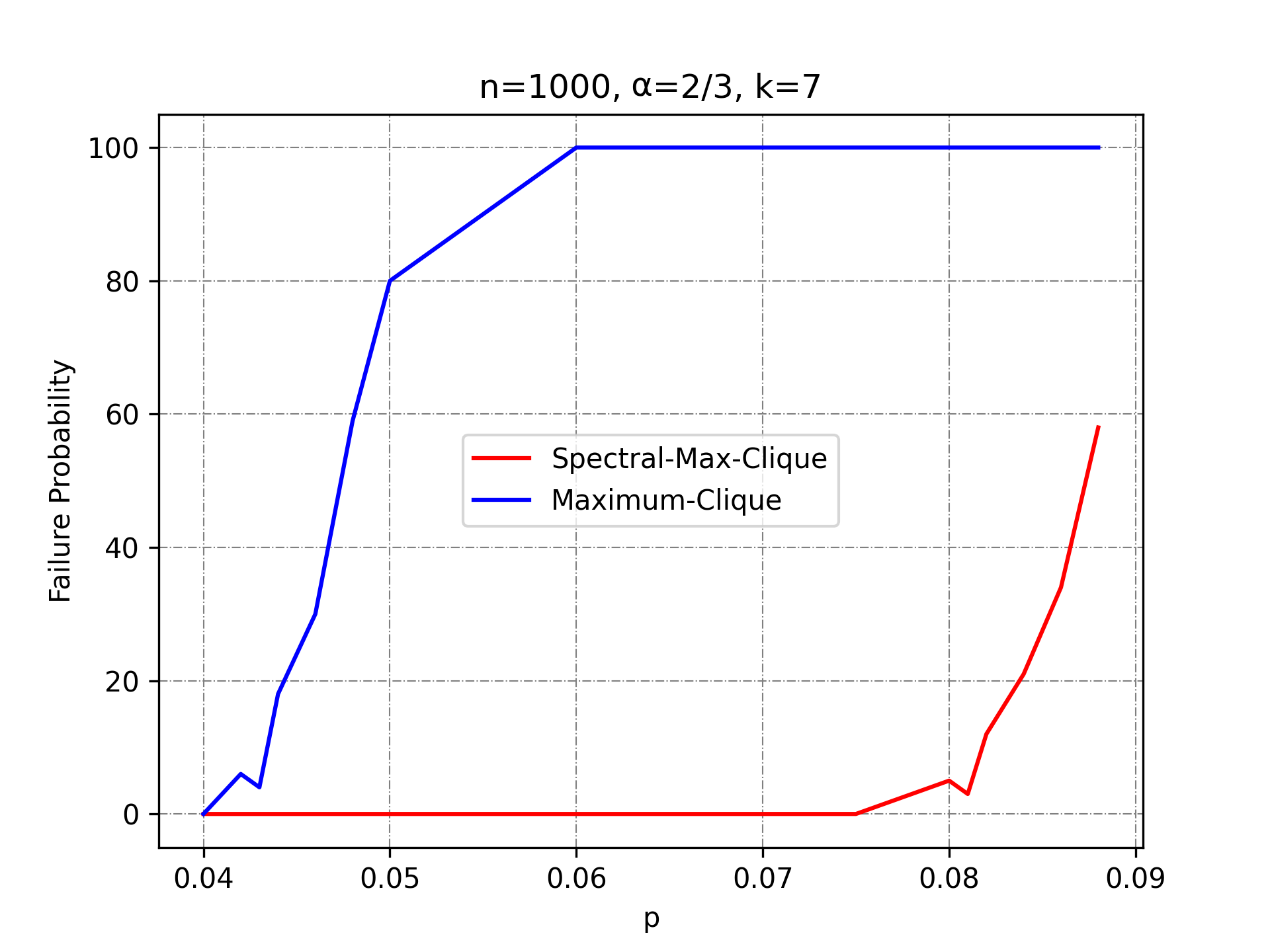}
        \caption{$k = 7$}\label{fig2b}
    \end{subfigure}
    
    \caption{Failure probability curves for $\alpha = 2/3$, $n = 1000$ and $k=4, 7$.}\label{fig2}
\end{figure*}

%\vspace{-1.5cm}
\begin{figure*}[ht]
    \centering
    \begin{subfigure}[b]{0.482\textwidth}
        \centering
        \includegraphics[width=\textwidth]{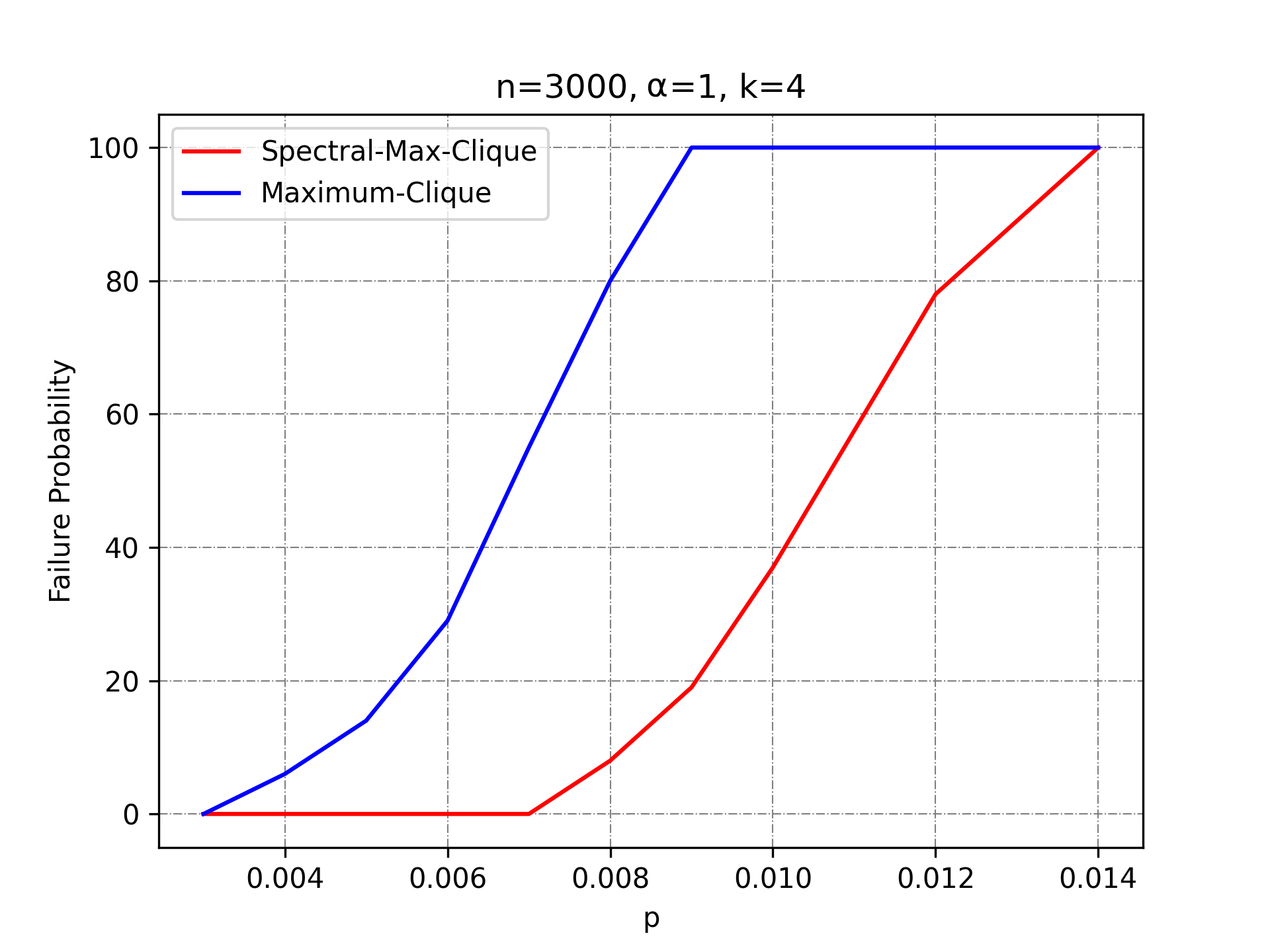}
        \caption{$k = 4$}\label{fig3a}
    \end{subfigure}
    \quad
    \begin{subfigure}[b]{0.482\textwidth}  
        \centering 
        \includegraphics[width=\textwidth]{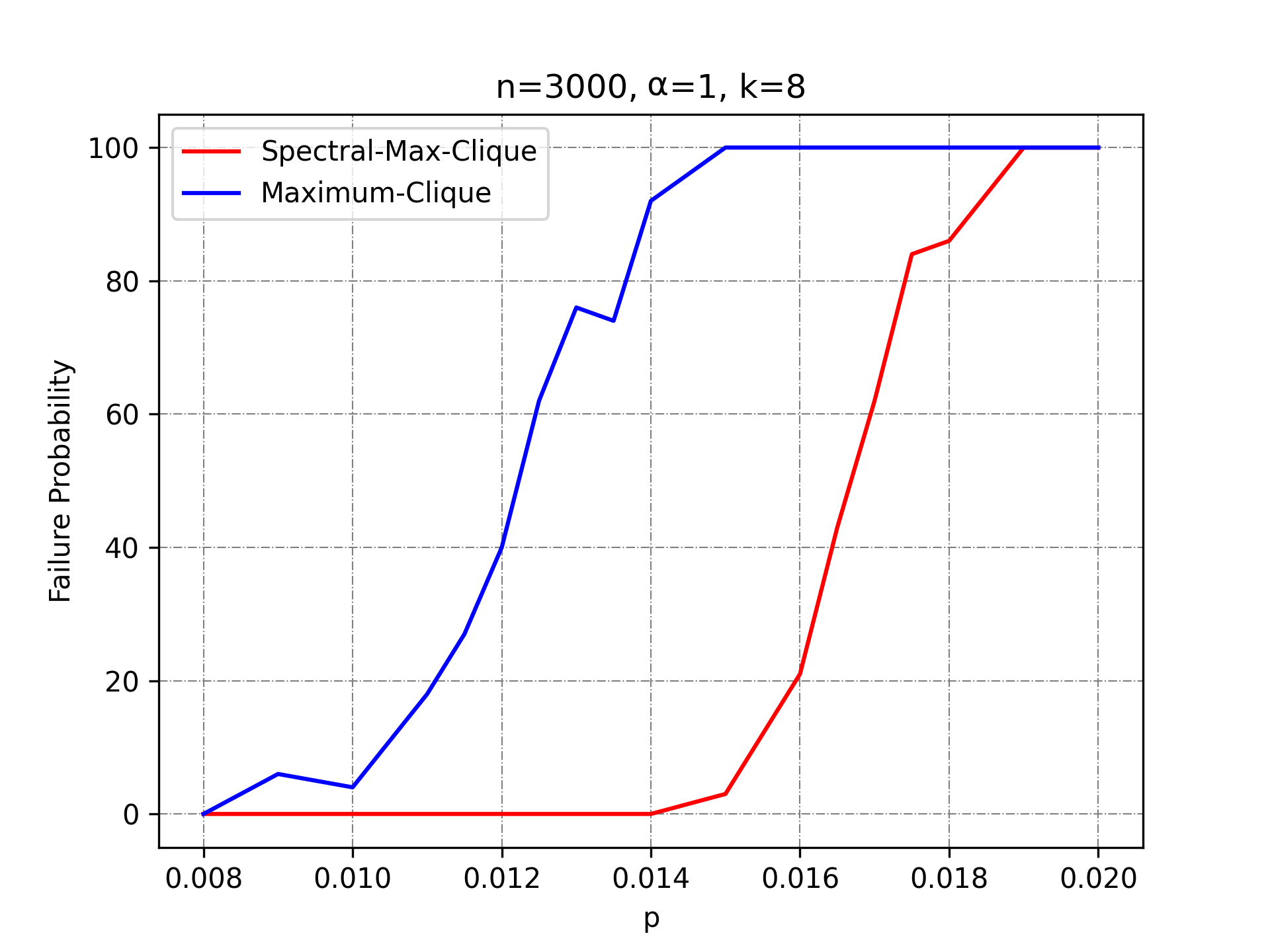}
        \caption{$k = 8$}\label{fig3b}
    \end{subfigure}
    
    \caption{Failure probability curves for $\alpha = 1$, $n = 3000$ and $k=4, 8$.}\label{fig3}
\end{figure*}

%\vspace{-0.5cm}
The \cref{fig1,,fig2,,fig3} show the failure probability of each algorithm, when $p$ increases, meaning that the $G_{n,m,p}$ becomes denser. These experiments show that for smaller values of $k$ and $p$, the two algorithms perform in a similar manner; they both find successfully the maximum clique in the graph. This is true for all the different values of parameter $a$. Nevertheless, for the case $a = 1$ and $k = 1$, fig. \ref{fig6a}, where the graph is relatively sparse, \emph{Spectral-Max-Clique} algorithm fails to find the maximum clique but strictly improves when $k$ gets larger or when the graph is more dense. Indeed, as it is demonstrated in fig. \ref{fig2b}, when $k = 7$, the \emph{Spectral-Max-Clique} algorithm has failure probability close to $0$ for the smaller values of $p$, while \emph{Maximum-Clique} fails to find the maximum clique in almost all the cases, with failure probability close to $100\%$. One more example is fig. \ref{fig1b}, where the failure probability of \emph{Maximum-Clique} starts at $p \approx 0.17$ and increases as the graph gets denser, and fails in all of the cases to find the maximum clique when $p \approx 0.27$. On the other hand, the probability of failure of \emph{Spectral-Max-Clique} begins when $p \approx 0.25$ and fails in all cases when $p \approx 0.36$. From all the figures it is obvious that the failure probability of the spectral algorithm also increases but slower than the failure probability of \emph{Maximum-Clique}.

%\vspace{-0.5cm}

\begin{figure*}[ht]
    \centering
    \begin{subfigure}[b]{0.482\textwidth}
        \centering
        \includegraphics[width=\textwidth]{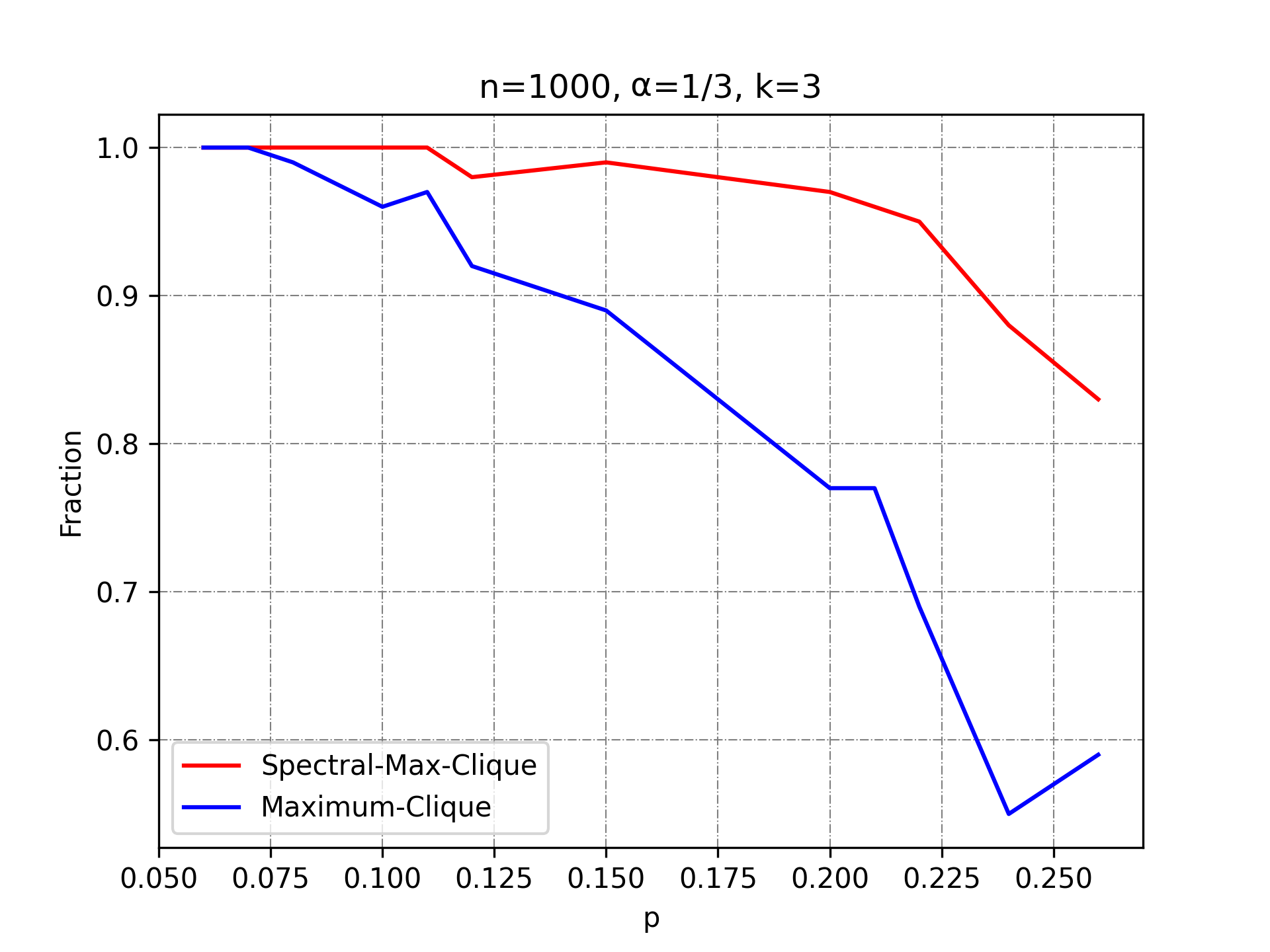}
        \caption{$k = 3$}\label{frac1a}
    \end{subfigure}
    \quad
    \begin{subfigure}[b]{0.482\textwidth}  
        \centering 
        \includegraphics[width=\textwidth]{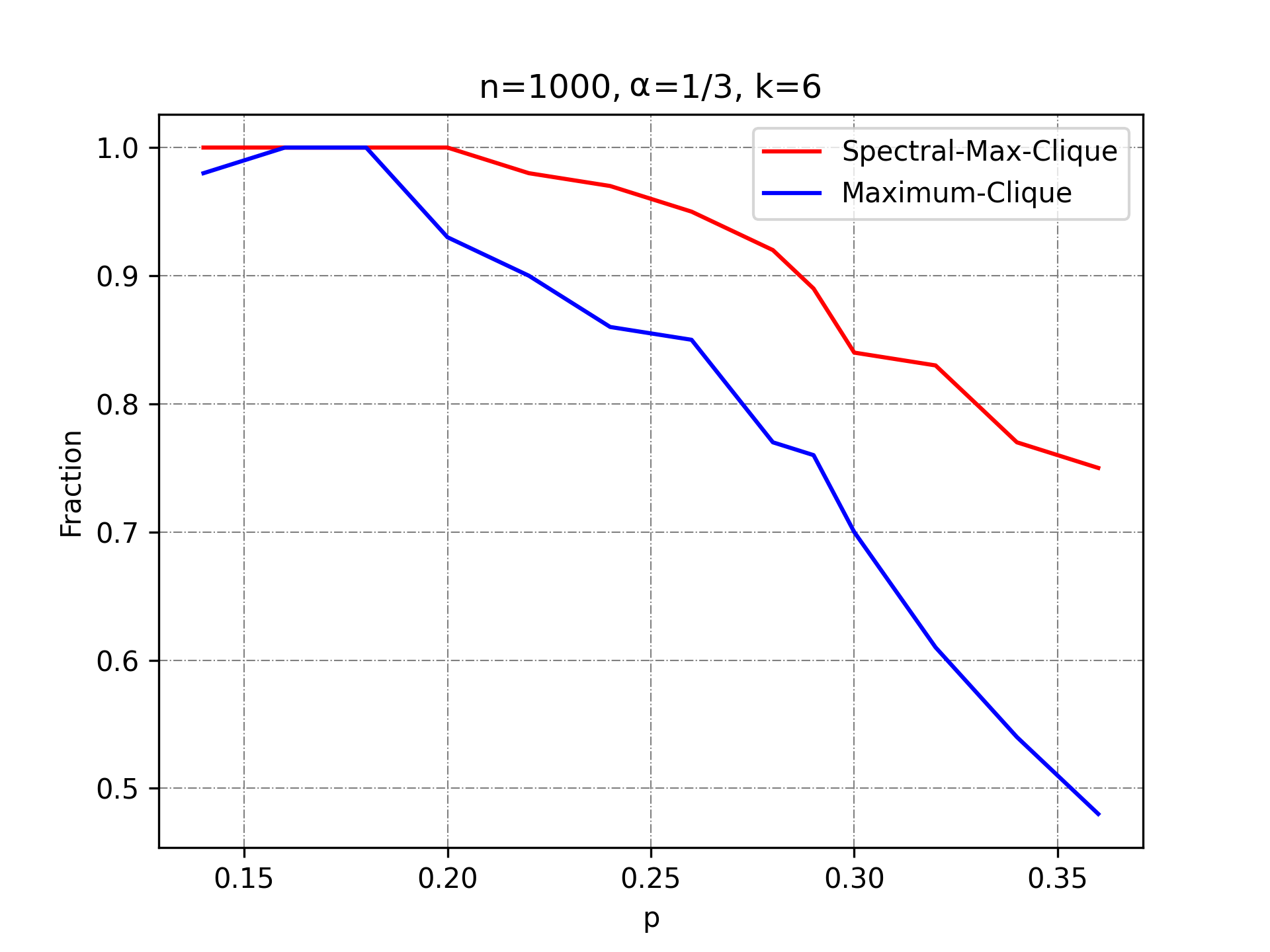}
        \caption{$k = 6$}\label{frac1b}
    \end{subfigure}
    
    \caption{Approximation guarantee curves for $\alpha = 1/3$, $n = 1000$ and $k=3,6$.}\label{frac1}
\end{figure*}

%\vspace{-0.8cm}

\begin{figure*}[ht]
    \centering
    \begin{subfigure}[b]{0.482\textwidth}
        \centering
        \includegraphics[width=\textwidth]{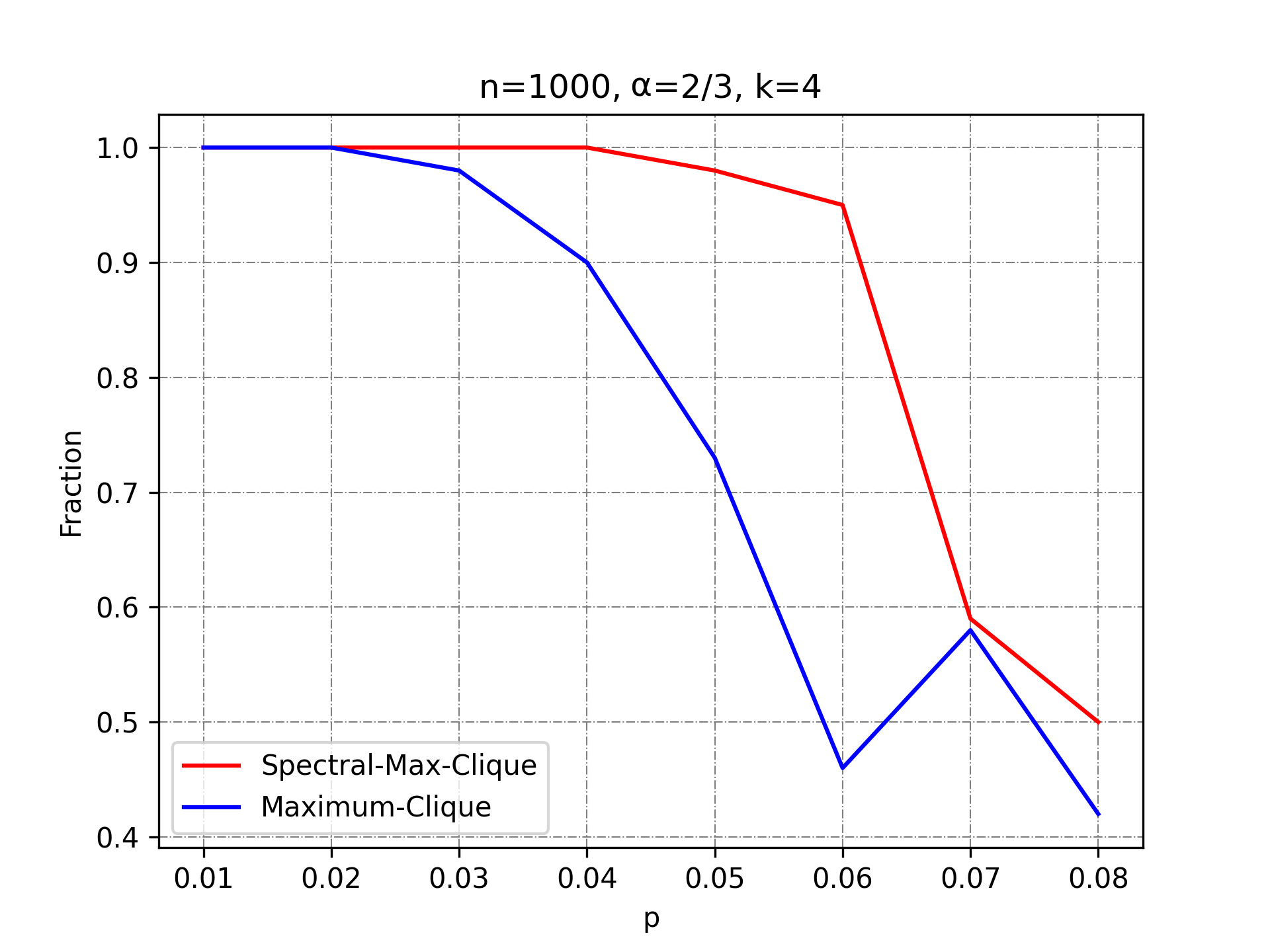}
        \caption{$k = 4$}\label{frac2a}
    \end{subfigure}
    \quad
    \begin{subfigure}[b]{0.482\textwidth}  
        \centering 
        \includegraphics[width=\textwidth]{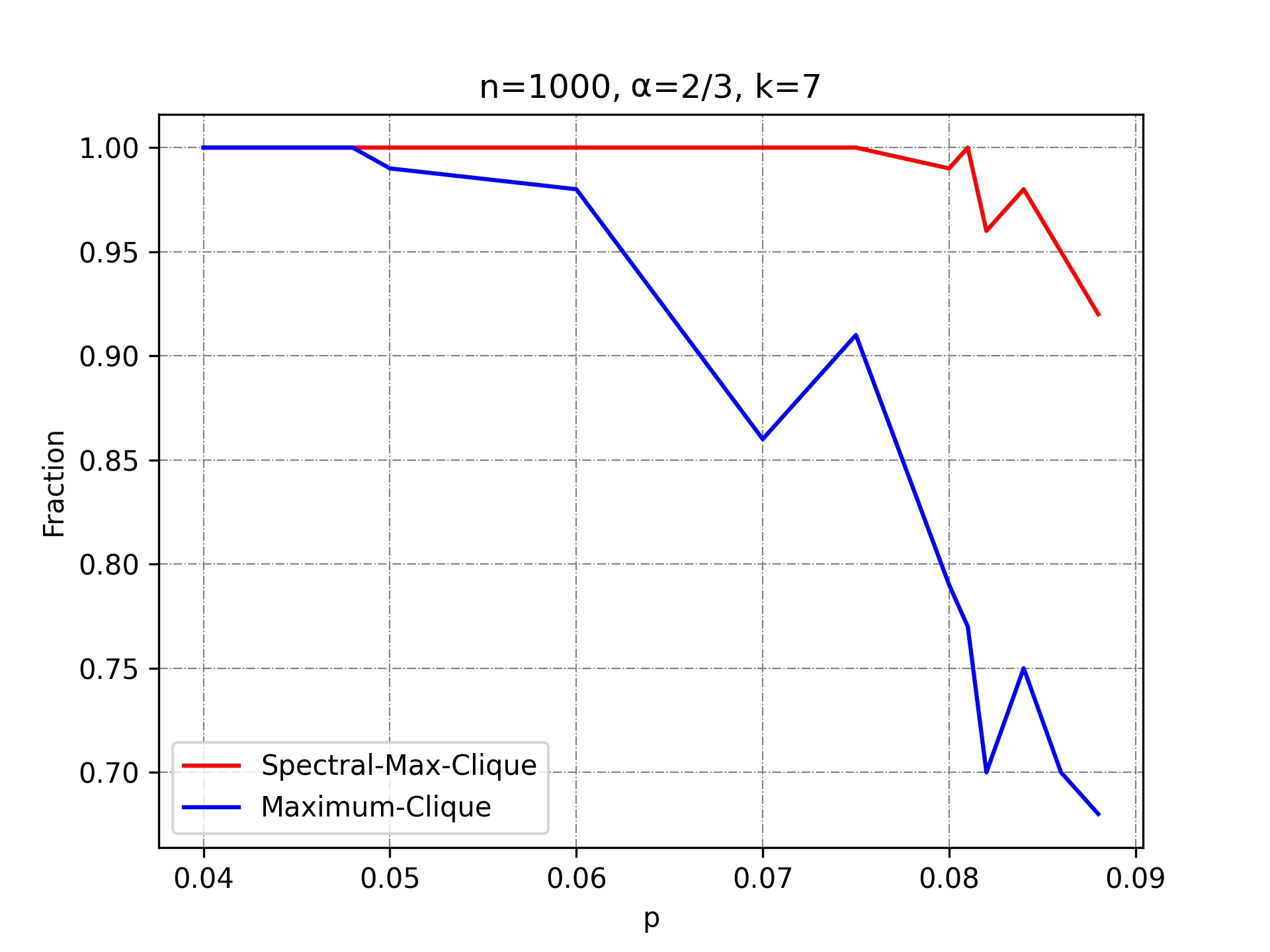}
        \caption{$k = 7$}\label{frac2b}
    \end{subfigure}
    
    \caption{Approximation guarantee curves for $\alpha = 2/3$, $n = 1000$ and $k=4,7$.}\label{frac2}
\end{figure*}

\begin{figure*}[ht]
    \centering
    \begin{subfigure}[b]{0.482\textwidth}
        \centering
        \includegraphics[width=\textwidth]{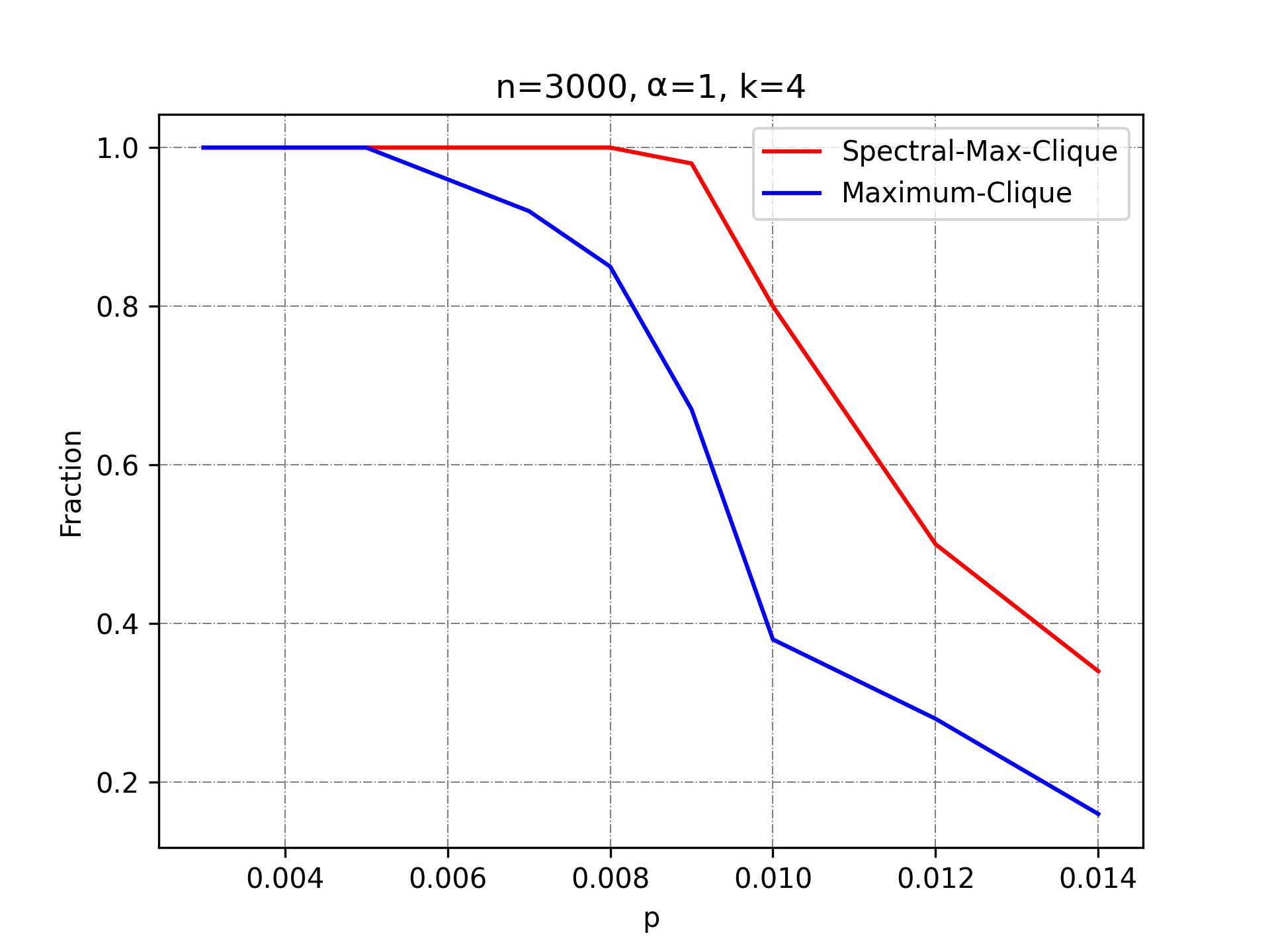}
        \caption{$k = 4$}\label{frac3a}
    \end{subfigure}
    \quad
    \begin{subfigure}[b]{0.482\textwidth}  
        \centering 
        \includegraphics[width=\textwidth]{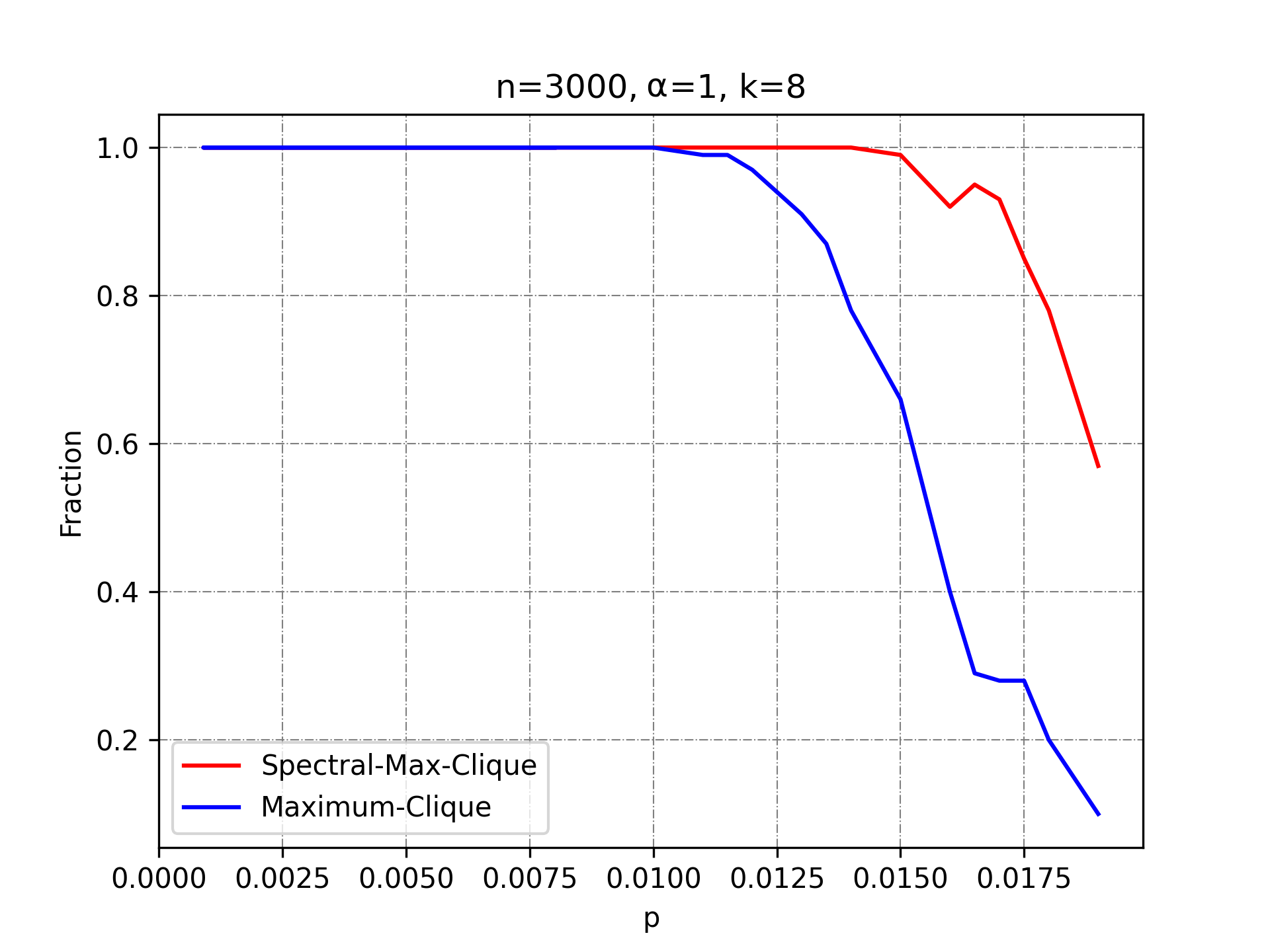}
        \caption{$k = 8$}\label{frac3b}
    \end{subfigure}
    
    \caption{Approximation guarantee curves for $\alpha = 1$, $n = 3000$ and $k=4,8$.}\label{frac3}
\end{figure*}
%\vspace{-0.5cm}

It is also interesting to demonstrate how far the resulting clique of each algorithm is from the maximum clique of the graph. For that reason, we ran experiments in the cases where both algorithms fail. In particular, \cref{frac1,,frac2,,frac3} show the curves of the average of the fraction of the clique size found by the algorithms over the maximum clique size of the input graph. By studying these figures, we can observe that for all the different values of parameters $\alpha$ and $k$, the output clique of \emph{Spectral-Max-Clique} algorithm is closer to the maximum clique of the graph with respect to the output clique of \emph{Maximum-Clique} algorithm; the size of the clique that the \emph{Spectral-Max-Clique} algorithm finds is closer to the size of the maximum clique of the graph. For instance, for the case $a = 1/3$ and $k = 3$, fig. \ref{frac1a}, when $p \approx 0.250$ and the graph is denser, the approximation guarantee for \emph{Spectral-Max-Clique} is $fraction \approx 0.82$ while for \emph{Maximum-Clique} is $fraction \approx 0.57$. One more apparent example is fig. \ref{frac3b}, when $a = 1$ and $k = 8$. In this case, $fraction \approx 0.6$ for \emph{Spectral-Max-Clique} algorithm, although for \emph{Maximum-Clique}, $fraction \approx 0.15$. Therefore, the former algorithm finds more than half of the maximum clique while the latter fails to find approximately $85\%$ of the maximum clique of the graph.  

It should be noted that, as the value of $\alpha$ gets closer to $0$, we were only able to run our experiments for smaller values of $n$, because random graph instances are denser and choosing the right $k$-clique $S$ that leads to the maximum clique is more time consuming.

We conclude that for the cases when the graph gets more dense or when parameter $k$ gets larger, \emph{Spectral-Max-Clique} has a lower failure probability as well as it succeeds to find a larger portion of the maximum clique of the graph. The spectral algorithm performs better in dense instances, while the other algorithms for dense graphs do not perform well, meaning that they fail to find the maximum clique for each instance of the graph. Hence, spectral algorithm works for a larger interval of $p$ than the other algorithms.

\section{Conclusions}

In this paper, we considered the problem of finding maximum cliques when the input graph is a random instance of the random intersection graphs model. Current algorithms for this problem are successful with high probability only for relatively sparse instances, leaving the dense case mostly unexplored.
We presented a spectral algorithm for finding large cliques that processes vertices according to respective values in the second largest eigenvector of the adjacency matrix of induced subgraphs of the input graph corresponding to common \linebreak neighbors of small cliques. Our experimental evaluation showed that our spectral algorithm clearly outperforms existing polynomial time algorithms, especially in the dense regime. A precise characterization of the performance guarantees of our algorithm using formal methods remains open for future work.
We believe that spectral properties of random intersection graphs may be also used to construct efficient algorithms for other NP-hard graph theoretical problems as well.  

%
% ---- Bibliography ----
%
% BibTeX users should specify bibliography style 'splncs04'.
% References will then be sorted and formatted in the correct style.
%
\bibliographystyle{splncs04}
\bibliography{SOFSEM22_-_MaxCliques}

\begin{thebibliography}{10}
\providecommand{\url}[1]{\texttt{#1}}
\providecommand{\urlprefix}{URL }
\providecommand{\doi}[1]{https://doi.org/#1}

\bibitem{AKS98}
Alon, N., Krivelevich, M., Sudakov, B.: Finding a large hidden clique in a
  random graph. Random Struct. Algor.  \textbf{13},  457--466 (1998)

\bibitem{BT06}
Behrisch, M., Taraz, A.: Efficiently covering complex networks with cliques of
  similar vertices. Theor. Comput. Sci.  \textbf{355}(1),  37--47 (2006)

\bibitem{survey2}
Bloznelis, M., Godehardt, E., Jaworski, J., Kurauskas, V., Rybarczyk, K.:
  Properties of random intersection graphs: Models of random intersection
  graphs. In: Data Science, Learning by Latent Structures, and Knowledge
  Discovery - revised versions of selected papers presented during the European
  Conference on Data Analysis (ECDA 2013), pp. 79--88. Springer (2015)

\bibitem{survey1}
Bloznelis, M., Godehardt, E., Jaworski, J., Kurauskas, V., Rybarczyk, K.:
  Recent progress in complex network analysis: Models of random intersection
  graphs. In: Data Science, Learning by Latent Structures, and Knowledge
  Discovery - revised versions of selected papers presented during the European
  Conference on Data Analysis (ECDA 2013), pp. 69--78. Springer (2015)

\bibitem{BK17}
Bloznelis, M., Kurauskas, V.: Large cliques in sparse random intersection
  graphs. Electr. J. Comb.  \textbf{24}(2), ~P2.5 (2017)

\bibitem{FSS00}
Fill, J.A., Sheinerman, E.R., Singer-Cohen, K.B.: Random intersection graphs
  when $m = \omega(n)$: an equivalence theorem relating the evolution of the
  $g(n, m, p)$ and $g(n, p)$ models. Random Struct. Algor.  \textbf{16}(2),
  156--176 (2000)

\bibitem{FH15}
Friedrich, T., Hercher, C.: On the kernel size of clique cover reductions for
  random intersection graphs. J. Discrete Algorithms  \textbf{34},  128--136
  (2015)

\bibitem{GM75}
Grimmett, G.R., McDiarmid, C.: On coloring random graphs. Math. Proc. Cambridge
  Philos. Soc.  \textbf{77},  313--324 (1975)

\bibitem{H99}
H\r{a}stad, J.: Clique is hard to approximate within $n^{1-\varepsilon}$. Acta
  Mathematica  \textbf{182},  105--142 (1999)

\bibitem{CHKX06}
Jianer, C., Xiuzhen, H., Iyad, A.K., Ge, X.: Strong computational lower bounds
  via parameterized complexity. J. Comput. Syst. Sci.  \textbf{72}(8),
  1346--1367 (2006)

\bibitem{KSS99}
Karo\'{n}ski, M., Scheinerman, E.R., Singer-Cohen, K.B.: On random intersection
  graphs: the subgraph problem. Comb. Probab Comput.  \textbf{8},  131--159
  (1999)

\bibitem{K72}
Karp, R.M.: Reducibility among combinatorial problems. In: Complexity of
  computer computations, pp. 85--103. Plenum Press (1972)

\bibitem{K76}
Karp, R.M.: Probabilistic analysis of some combinatorial search problems. In:
  Algorithms and Complexity: New Directions and Recent Results, pp. 85--103.
  Academic Press (1976)

\bibitem{NRS11}
Nikoletseas, S.E., Raptopoulos, C.L., Spirakis, P.G.: Communication and
  security in random intersection graphs models. In: 12th IEEE International
  Symposium on a World of Wireless, Mobile and Multimedia Networks (WOWMOM).
  pp.~1--6 (2011)

\bibitem{NRS12}
Nikoletseas, S.E., Raptopoulos, C.L., Spirakis, P.G.: Maximum cliques in graphs
  with small intersection number and random intersection graphs. In:
  Proceedings of the 37th International Symposium on Mathematical Foundations
  of Computer Science (MFCS). pp. 728--739 (2012)

\bibitem{TCS21}
Nikoletseas, S.E., Raptopoulos, C.L., Spirakis, P.G.: Maximum cliques in graphs
  with small intersection number and random intersection graphs. Comput. Sci.
  Rev.  \textbf{39},  100353 (2021)

\bibitem{R11}
Rybarczyk, K.: Equivalence of a random intersection graph and $g(n, p)$. Random
  Struct. Algor.  \textbf{38}(1-2),  205--234 (2011)

\bibitem{S95}
Singer-Cohen, K.B.: Random intersection graphs. Ph.D. thesis, John Hopkins
  University (1995)

\end{thebibliography}

\section{Appendix}
\subsection{Greedy-Clique Algorithm} \label{sec:greedy-clique}

The pseudocode of the GREEDY-CLIQUE Algorithm from \cite{BK17} is shown below.

\begin{breakablealgorithm}
  \caption{GREEDY-CLIQUE \cite{BK17}}
  \begin{algorithmic}[1]
    \Input Random instance $G$ of ${\cal G}_{n,m,p}$\
    \Output Clique $Q$ \
    \State Let $v_1,\ldots, v_2$ the vertices of $G$ in order of decreasing degree;
    \State $Q = \emptyset;$ \ 
    \ForEach{$i=1$ to $n$}
        \If{$v_i$ is adjacent to each vertex in $Q$}
            \State $Q = Q \cup \text{\{} v_i \text{\}};$ \

        \EndIf
    \EndFor
    \State \Return $Q$
\end{algorithmic}
\end{breakablealgorithm}

\subsection{Mono-Clique Algorithm} \label{sec:mono-clique}

The pseudocode of the MONO-CLIQUE Algorithm from \cite{BK17} is shown below.

\begin{breakablealgorithm}
  \caption{MONO-CLIQUE \cite{BK17}}
  \begin{algorithmic}[1]
    \Input Random instance $G$ of ${\cal G}_{n,m,p}$\
    \Output Clique $Q$ \
    \ForEach{$\{u,v\} \in E(G)$}
        \State $D(\{u,v\}) = |N(u) \cap N(v)|;$
    \EndFor
    \ForEach{$\{u,v\} \in E(G)$ in order of decreasing $D(\{u,v\})$}
        \State $S = N(u) \cap N(v);$
        \If{$S$ is a clique}
            \State \Return $Q = S \cup \text{\{}u,v\text{\}}$
        \EndIf
    \EndFor
    \State \Return any vertex $v \in V(G)$
\end{algorithmic}
\end{breakablealgorithm}

\subsection{Maximum-Clique Algorithm} \label{sec:maximum-clique}

The pseudocode of the Maximum-Clique Algorithm from \cite{BT06} is shown below.

\begin{breakablealgorithm}
  \caption{Maximum-Clique \cite{BT06}}
  \begin{algorithmic}[1]
    \Input Random instance $G$ of ${\cal G}_{n,m,p}$ and (fixed) parameter $k \in \mathbb{N}$\
    \Output Clique $Q$ of $G$\
%    \State $T := \emptyset;$\
    \State ${\cal L} = \emptyset;$\
    \ForEach{$U_k = \text{\{}v_1,\ldots,v_k\text{\}} \subseteq V$ such that $G[U_k]$ is complete}
        %\State remove $k$-subsets that do not form a clique;\
        \If{$\exists L \in {\cal L}: U_k \subseteq L$}
            \State \textbf{continue} to the next $U_k;$\
        \EndIf
%        \State $L := \emptyset;$\
%        \ForEach{$U_k = \text{\{}v_1,\ldots,v_k\text{\}} \subseteq V$}
            \State Let $Z = Z(U_k) := \cap_{k=1}^{i} N(v_i);$\
%            \ForEach{$u_i$, adjacent to $v_i \in U_k$}
%                \If{$N(v_i) == N(u_i)$}
%                    \State $T = T + u_i;$
%                \EndIf
%            \EndFor
            \If{$G[Z]$ is complete}
                \State ${\cal L} = {\cal L} \cup \{Z\};$
            \EndIf
%        \EndFor
    \EndFor
    \State ${\cal M} = \emptyset$;\
    \State  $Y = \emptyset;$\
    \ForEach{$Z \in {\cal L}$ in decreasing order $|Z|$}
        \If{$E(G[Z]) \not\subseteq Y$}
            \State $Y = Y \cup E(G[Z]);$\
            \State ${\cal M} = {\cal M} \cup \{Z\};$\
        \EndIf
    \EndFor
    \State Sort ${\cal M}$ by decreasing order;\
    \State \Return  largest $Q \subset {\cal M}$\
\end{algorithmic}
\end{breakablealgorithm}

\subsection{Further Experiments} \label{sec:further-experiments}

\begin{figure*}[ht]
    \centering
    \begin{subfigure}[b]{0.482\textwidth}
        \centering
        \includegraphics[width=\textwidth]{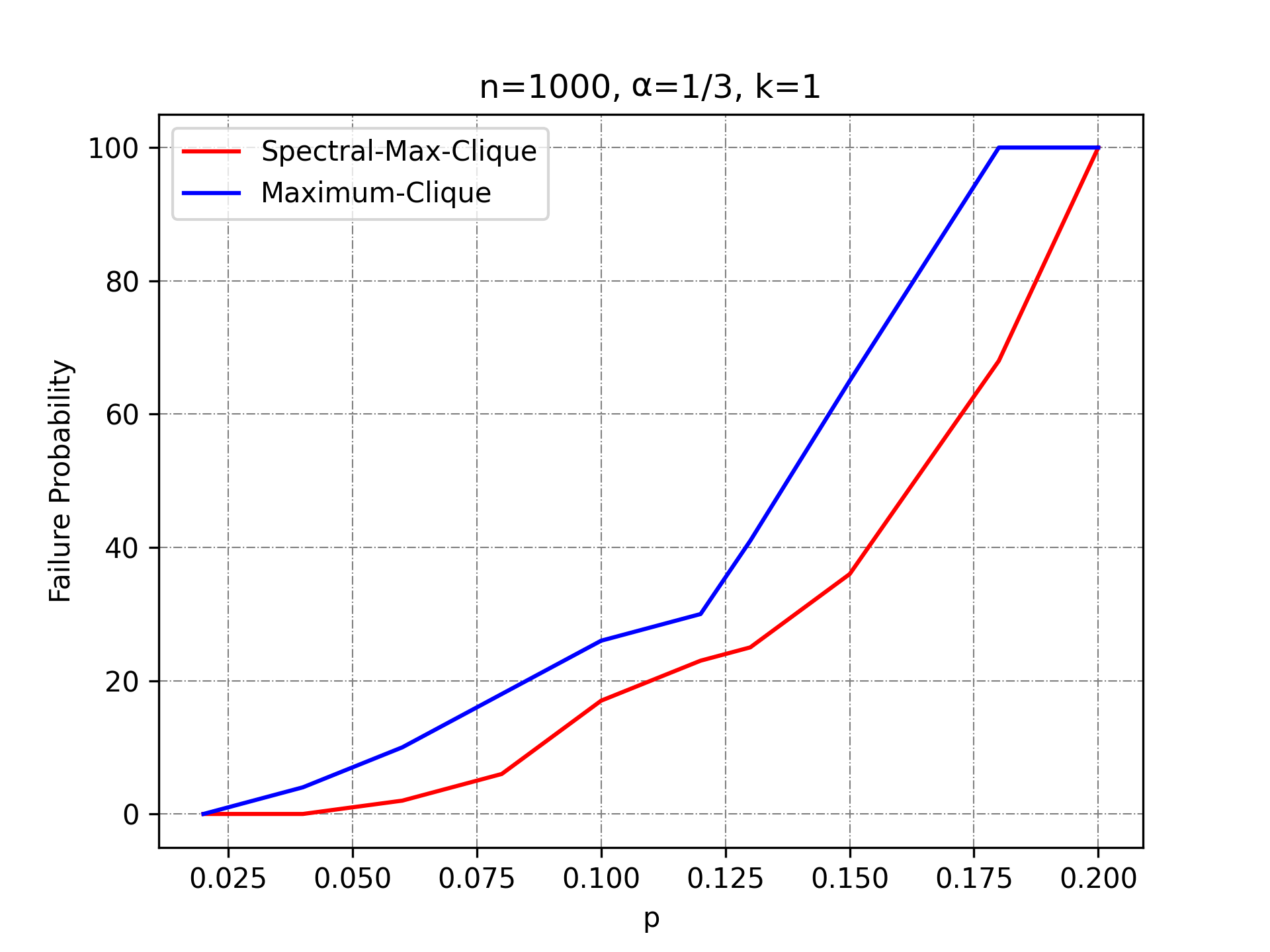}
        \caption{$k = 1$}
    \end{subfigure}
    \quad
    \begin{subfigure}[b]{0.482\textwidth}  
        \centering 
        \includegraphics[width=\textwidth]{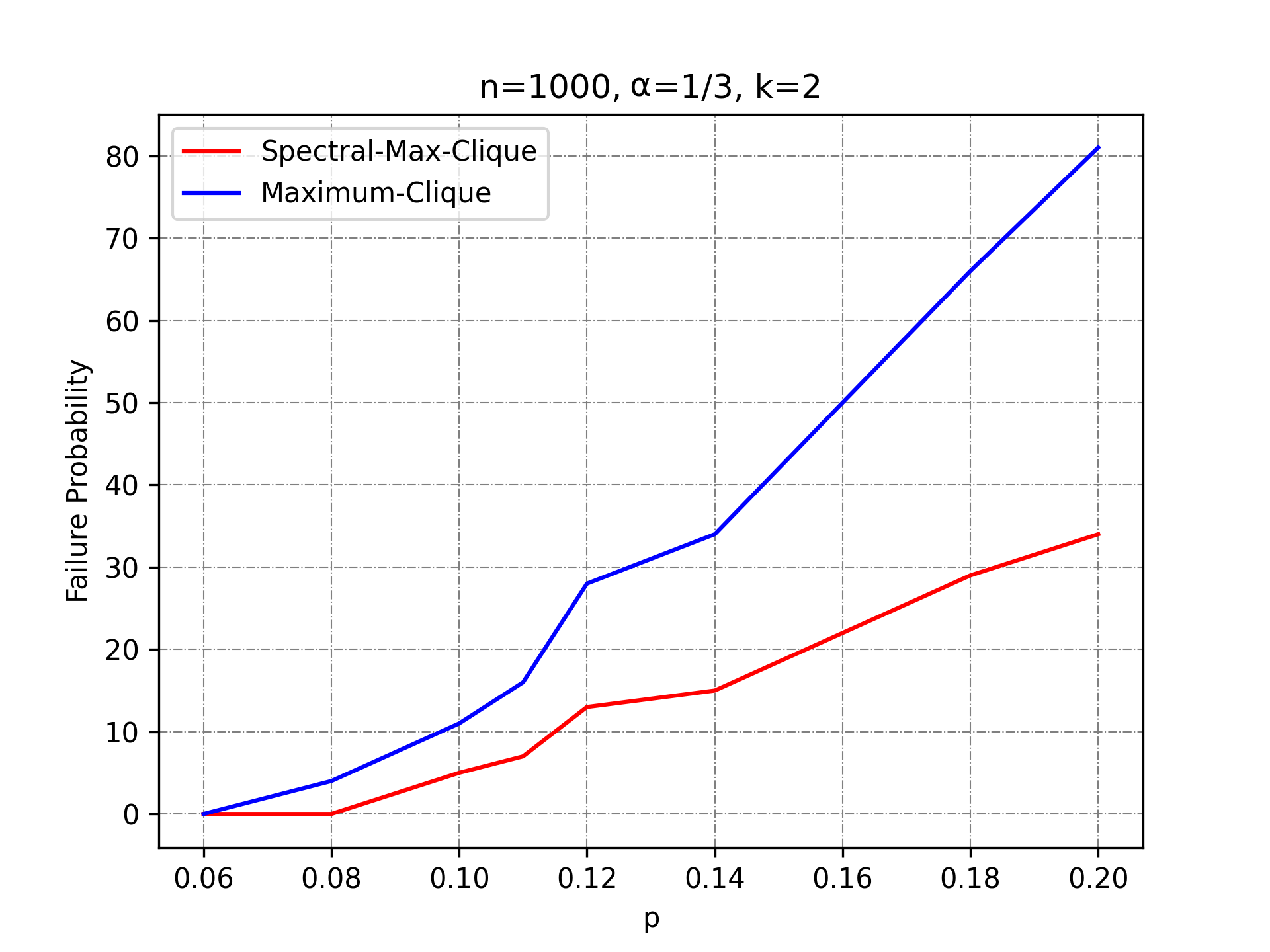}
        \caption{$k = 2$}
    \end{subfigure}
    \vskip\baselineskip
    \begin{subfigure}[b]{0.482\textwidth}   
        \centering 
        \includegraphics[width=\textwidth]{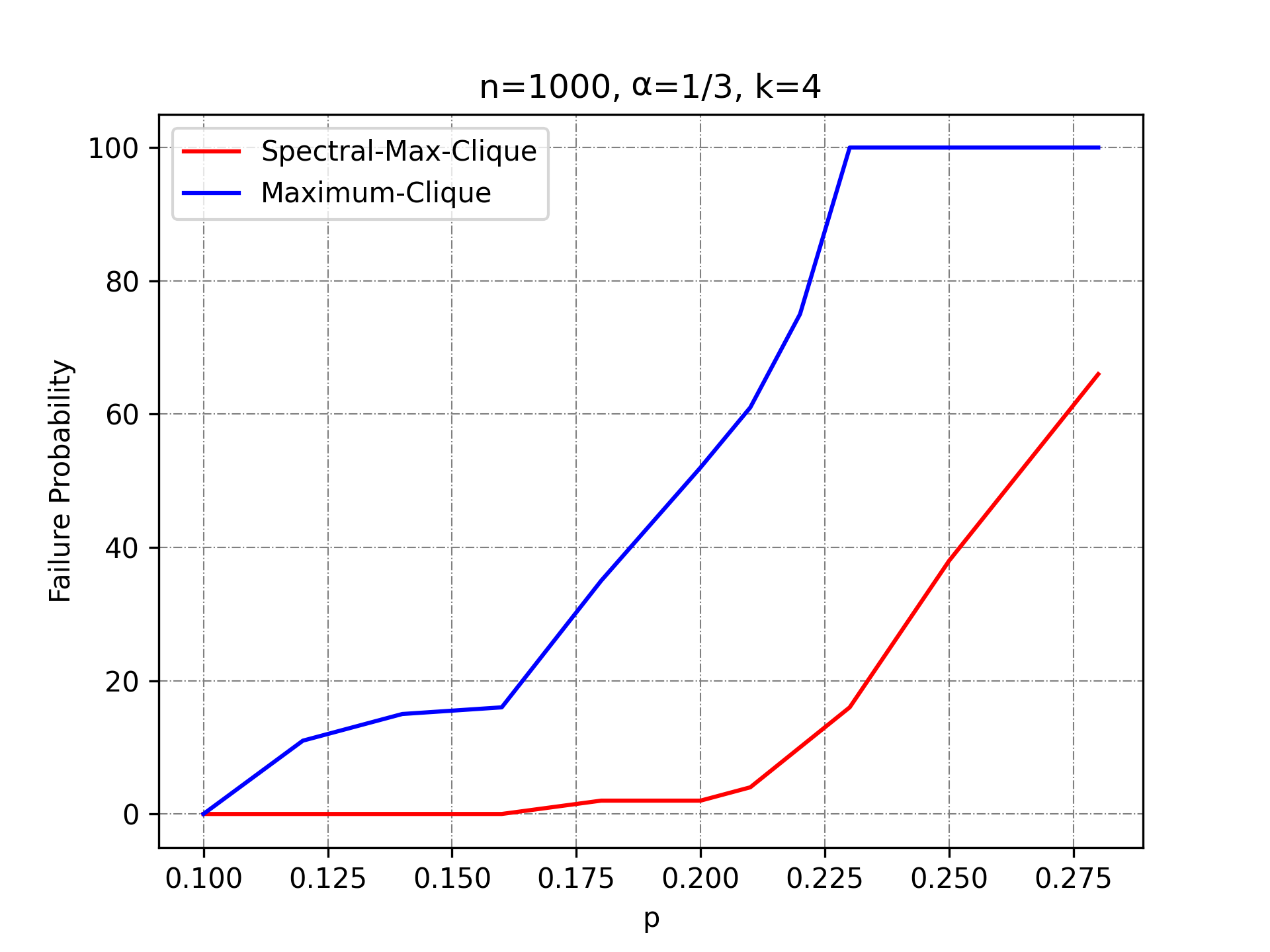}
        \caption{$k = 4$}
    \end{subfigure}
    \quad
    \begin{subfigure}[b]{0.482\textwidth}
        \centering
        \includegraphics[width=\textwidth]{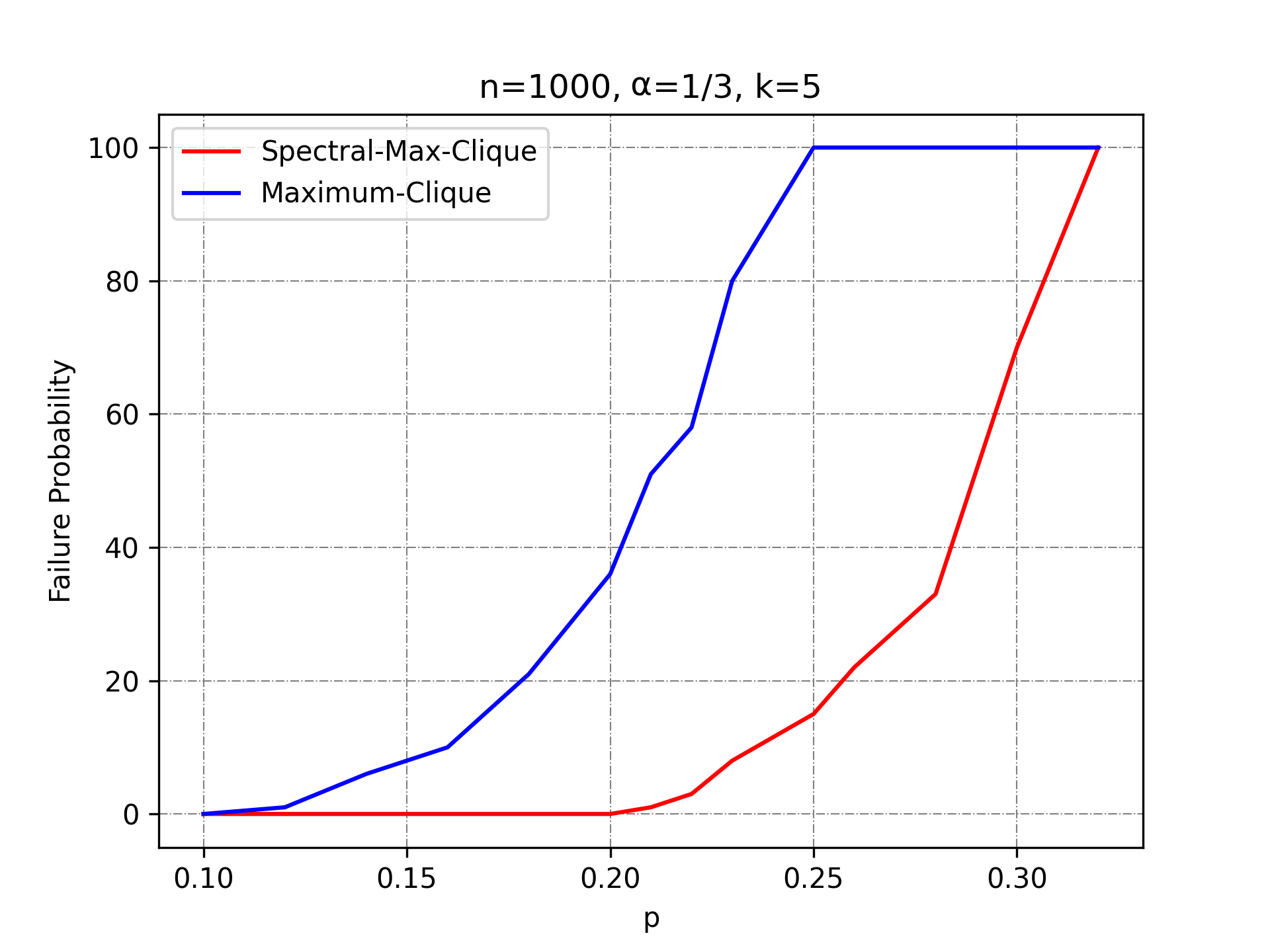}
        \caption{$k = 5$}
    \end{subfigure}
    \quad
    \caption{Failure probability curves for $\alpha = 1/3$, $n = 1000$ and $k=1, 2, 4, 5$.}\label{fig4}
\end{figure*}

\begin{figure*}[ht!]
    \centering
    \begin{subfigure}[b]{0.482\textwidth}
        \centering
        \includegraphics[width=\textwidth]{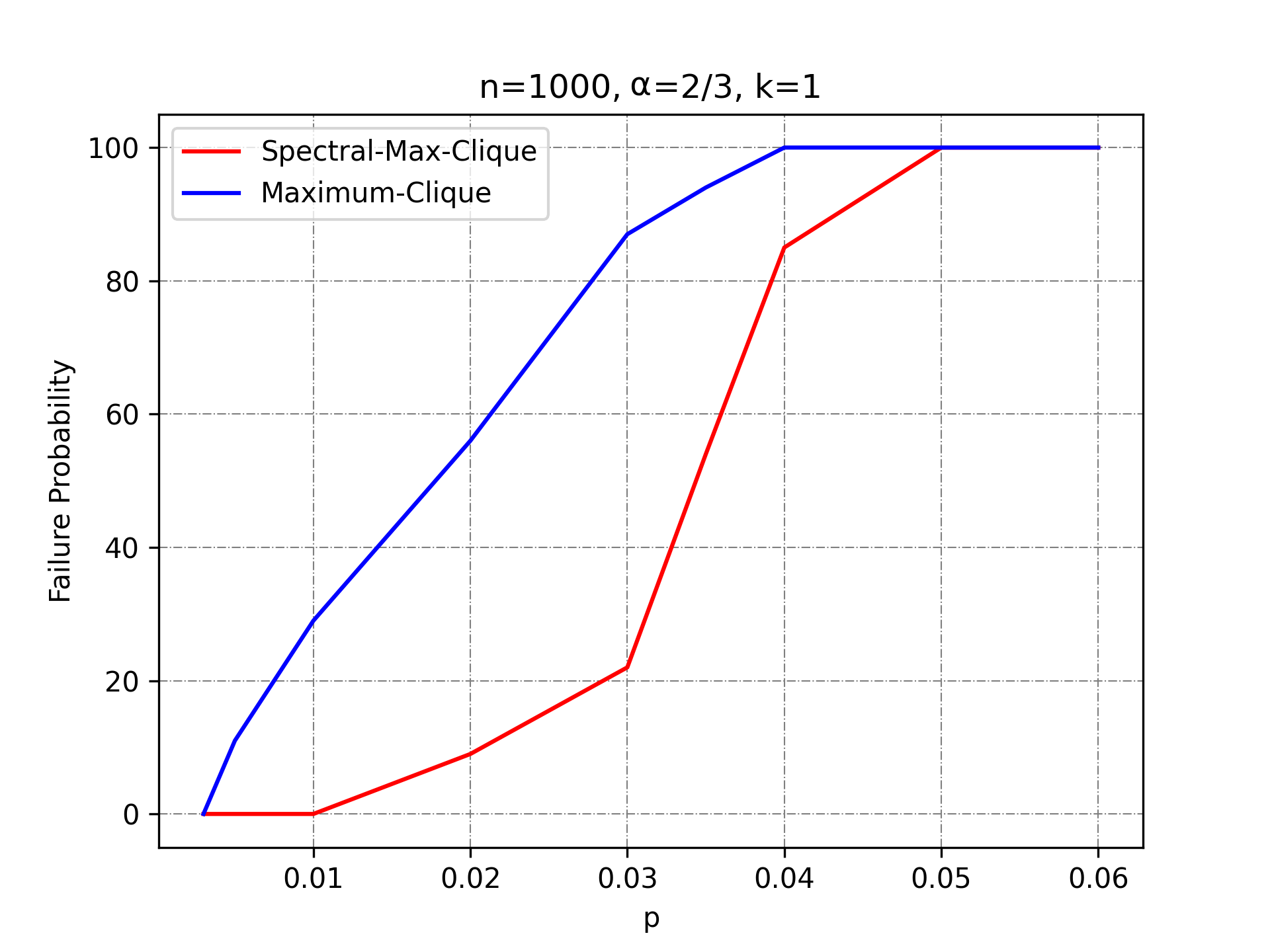}
        \caption{$k = 1$}
    \end{subfigure}
    \quad
    \begin{subfigure}[b]{0.482\textwidth}  
        \centering 
        \includegraphics[width=\textwidth]{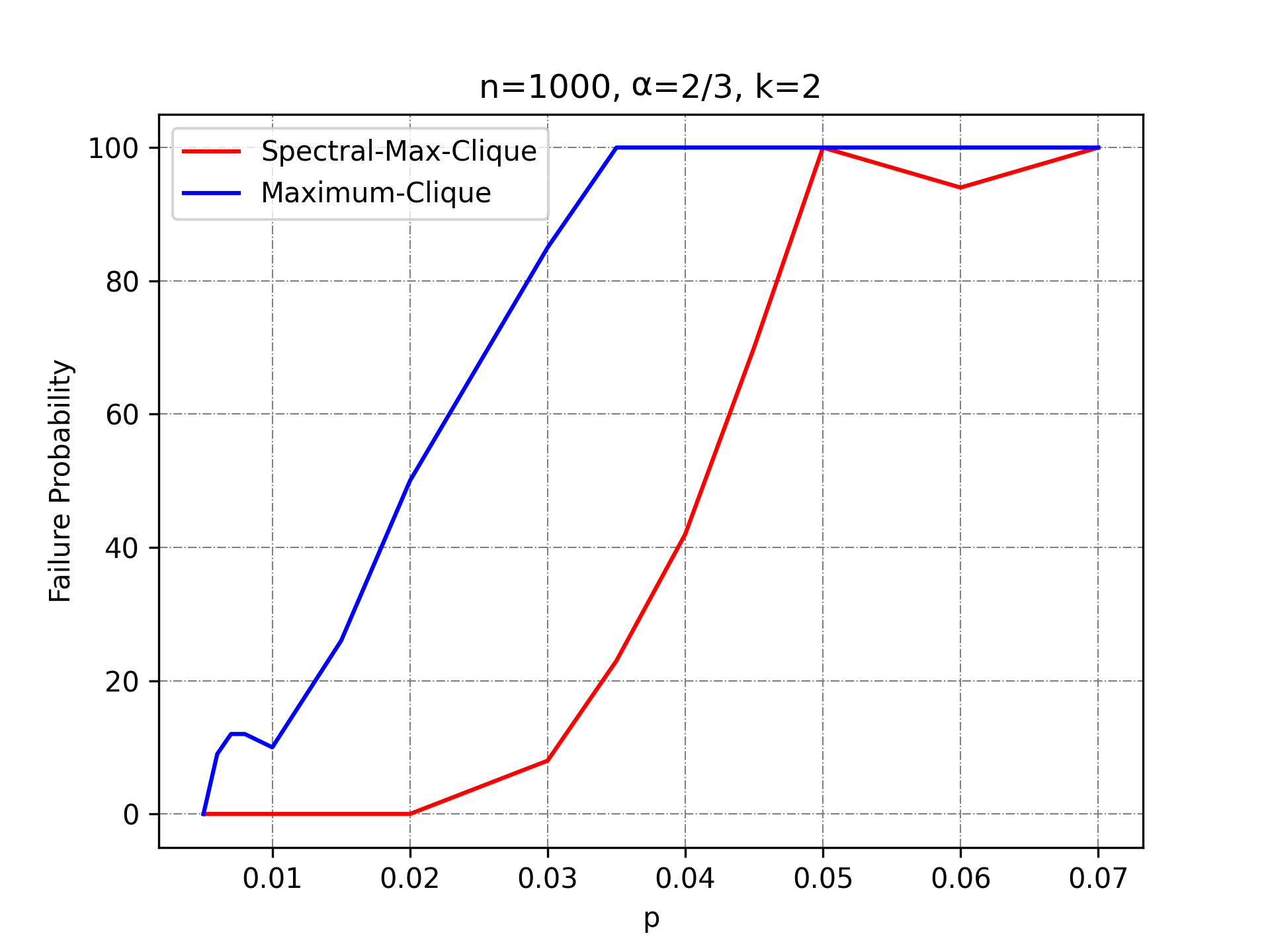}
        \caption{$k = 2$}
    \end{subfigure}
    
    \vskip\baselineskip
    \begin{subfigure}[b]{0.482\textwidth}   
        \centering 
        \includegraphics[width=\textwidth]{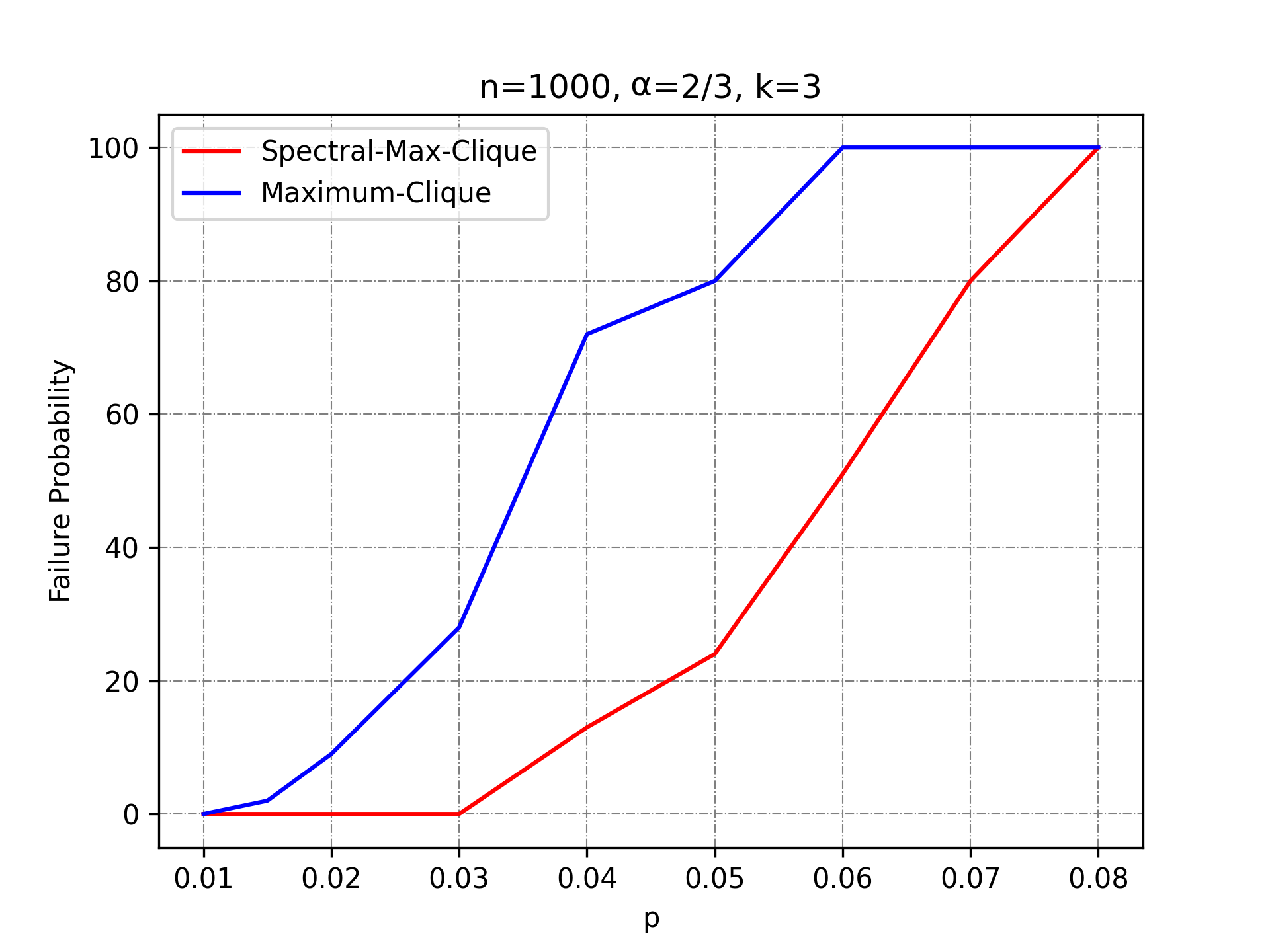}
        \caption{$k = 3$}
    \end{subfigure}
    \quad
    \begin{subfigure}[b]{0.482\textwidth}   
        \centering 
        \includegraphics[width=\textwidth]{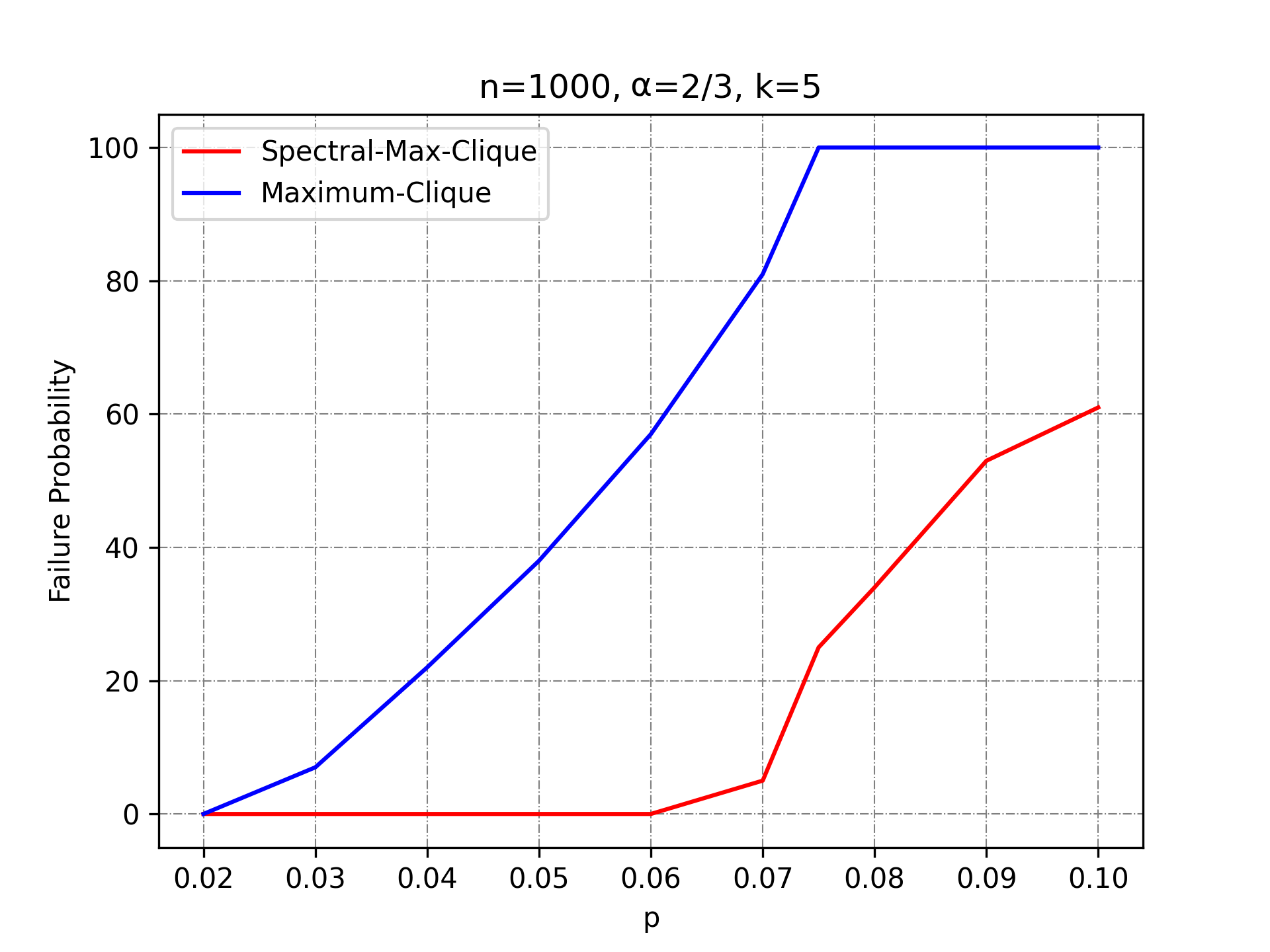}
        \caption{$k = 5$}
    \end{subfigure}
    \vskip\baselineskip
    \begin{subfigure}[b]{0.482\textwidth}   
        \centering 
        \includegraphics[width=\textwidth]{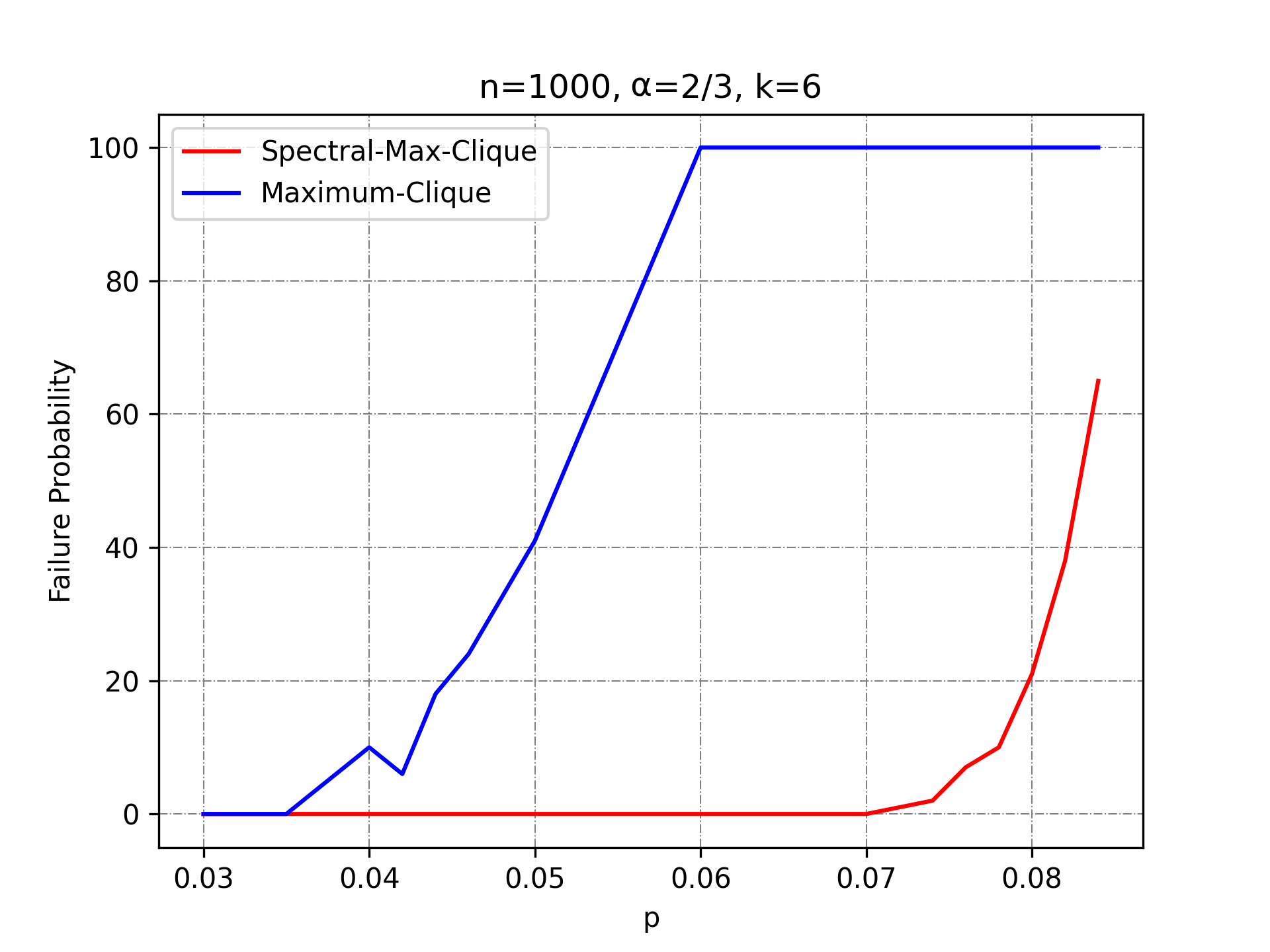}
        \caption{$k = 6$}
    \end{subfigure}
    
    \caption{Failure probability curves for $\alpha = 2/3$, $n = 1000$ and $k=1, 2, 3, 5, 6$.}\label{fig5}
\end{figure*}

\begin{figure*}[ht!]
    \centering
    \begin{subfigure}[b]{0.482\textwidth}
        \centering
        \includegraphics[width=\textwidth]{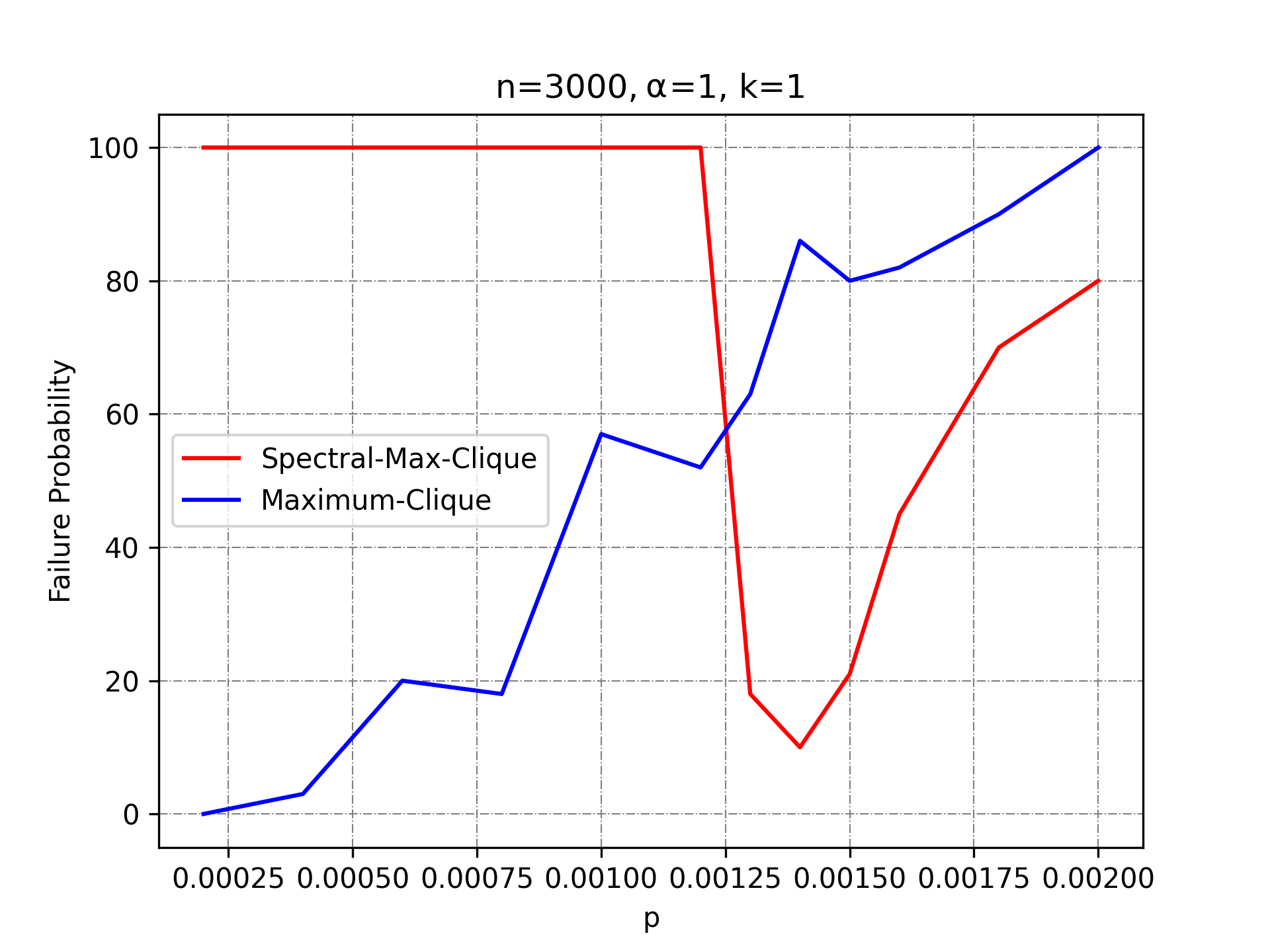}
        \caption{$k = 1$}\label{fig6a}
    \end{subfigure}
    \quad
    \begin{subfigure}[b]{0.482\textwidth}  
        \centering 
        \includegraphics[width=\textwidth]{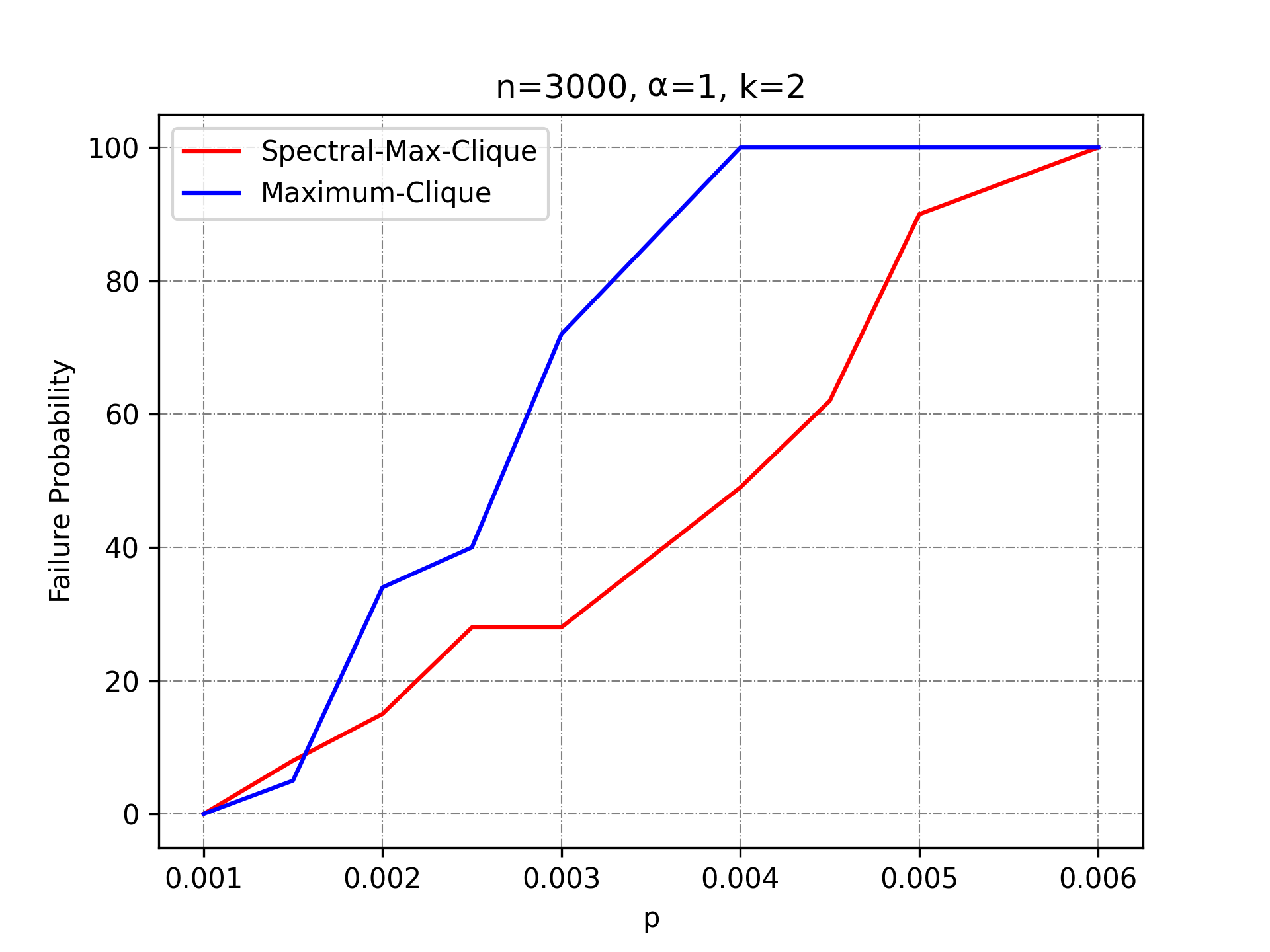}
        \caption{$k = 2$}
    \end{subfigure}
    
    \vskip\baselineskip
    \begin{subfigure}[b]{0.482\textwidth}   
        \centering 
        \includegraphics[width=\textwidth]{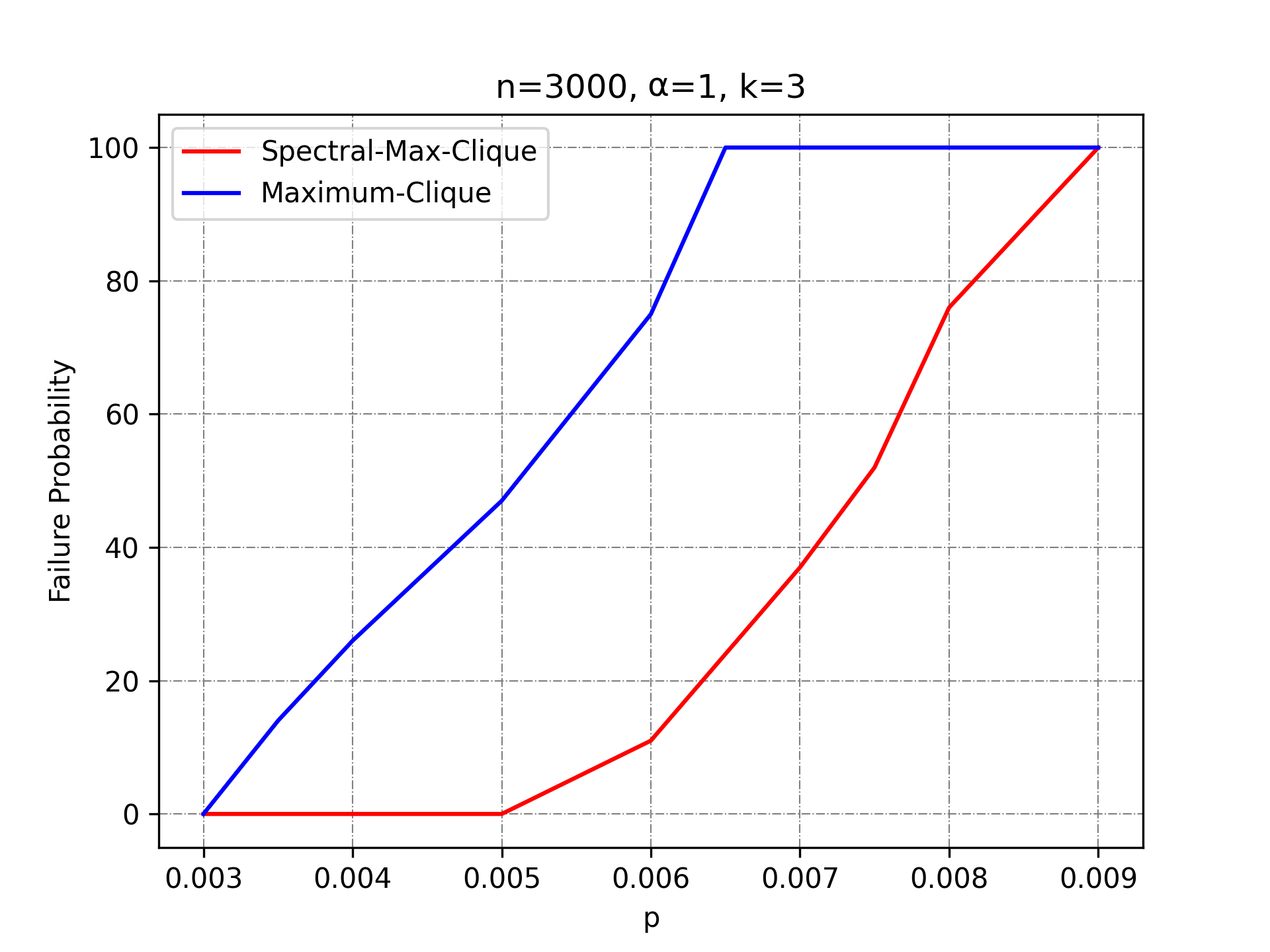}
        \caption{$k = 3$}
    \end{subfigure}
    \quad
    \begin{subfigure}[b]{0.482\textwidth}   
        \centering 
        \includegraphics[width=\textwidth]{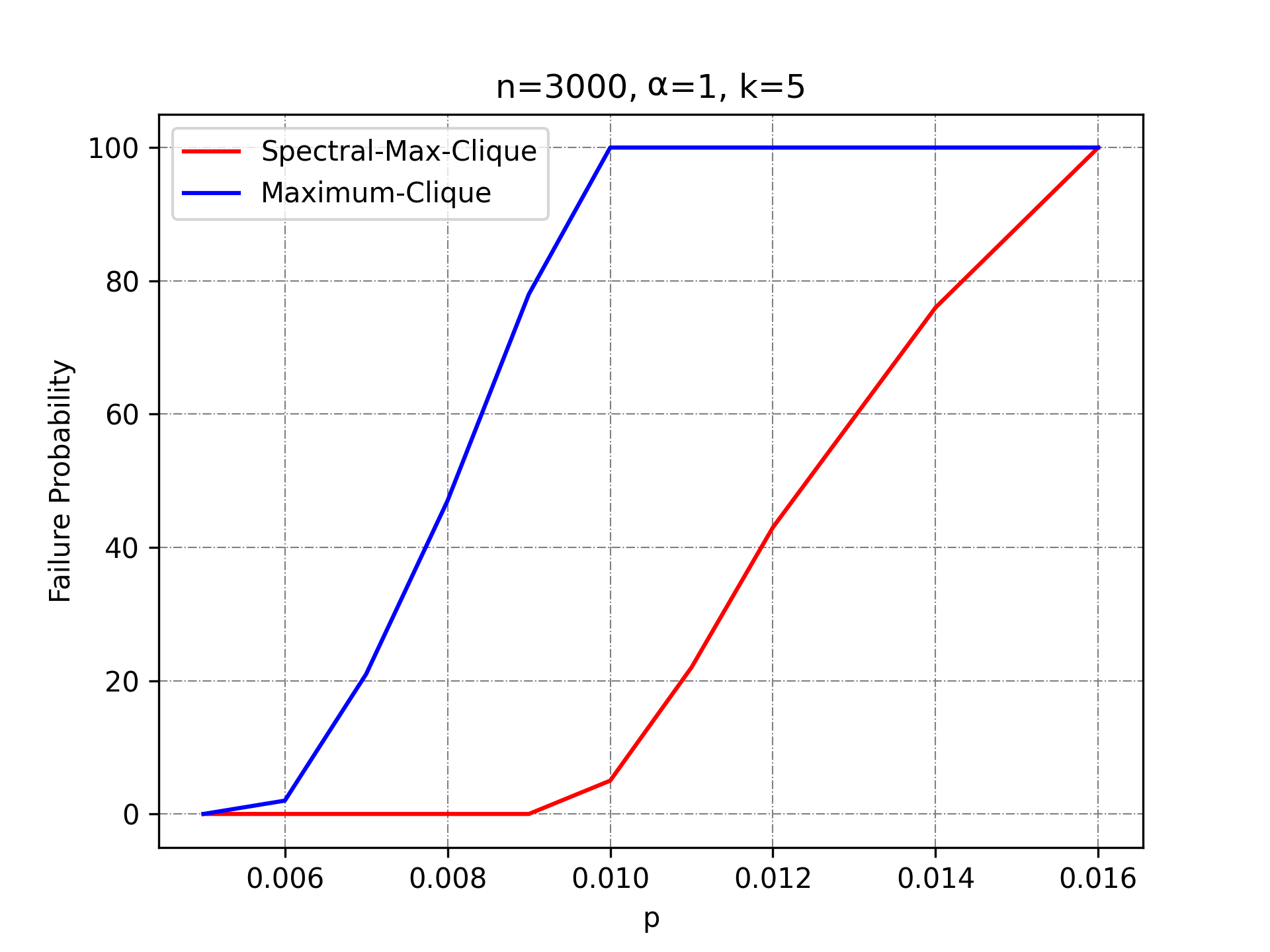}
        \caption{$k = 5$}
    \end{subfigure}
    
    \vskip\baselineskip
    \begin{subfigure}[b]{0.482\textwidth}   
        \centering 
        \includegraphics[width=\textwidth]{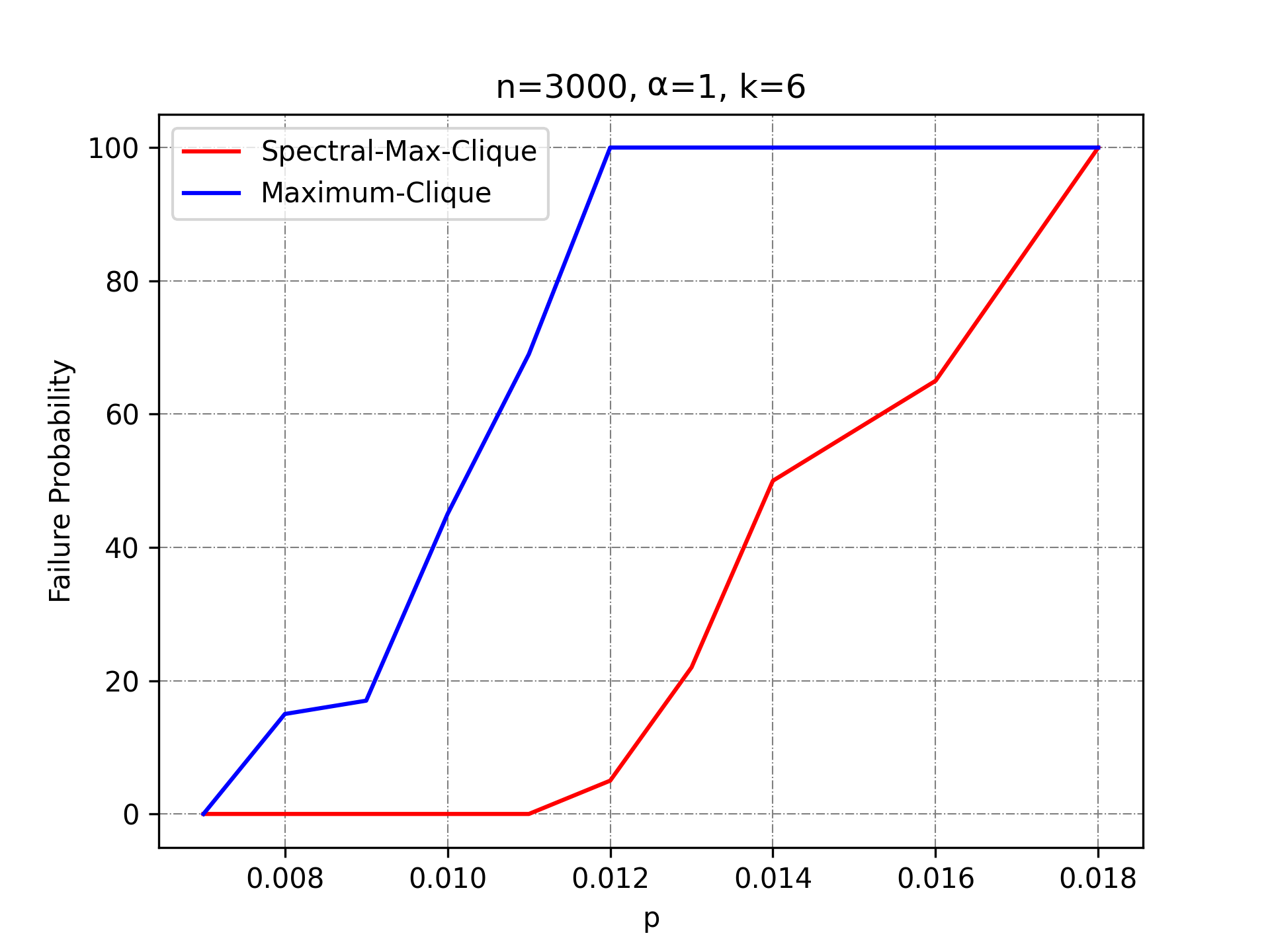}
        \caption{$k = 6$}
    \end{subfigure}
    \quad
    \begin{subfigure}[b]{0.482\textwidth}   
        \centering 
        \includegraphics[width=\textwidth]{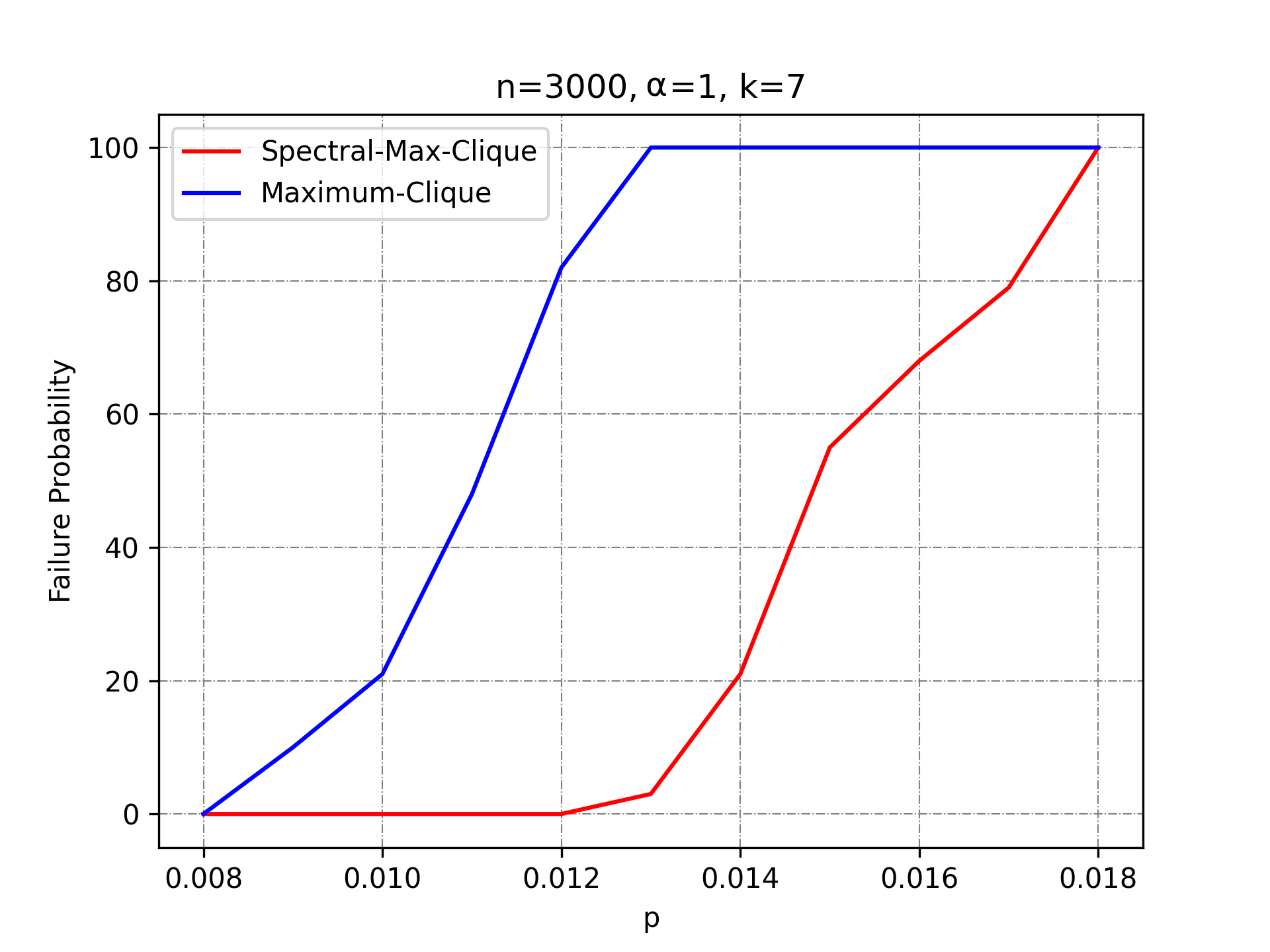}
        \caption{$k = 7$}
    \end{subfigure}
    
    \caption{Failure probability curves for $\alpha = 1$, $n = 3000$ and $k=1, 2, 3, 5, 6, 7$.}\label{fig6}
    
\end{figure*}

\begin{figure*}[ht]
    \centering
    \begin{subfigure}[b]{0.482\textwidth}
        \centering
        \includegraphics[width=\textwidth]{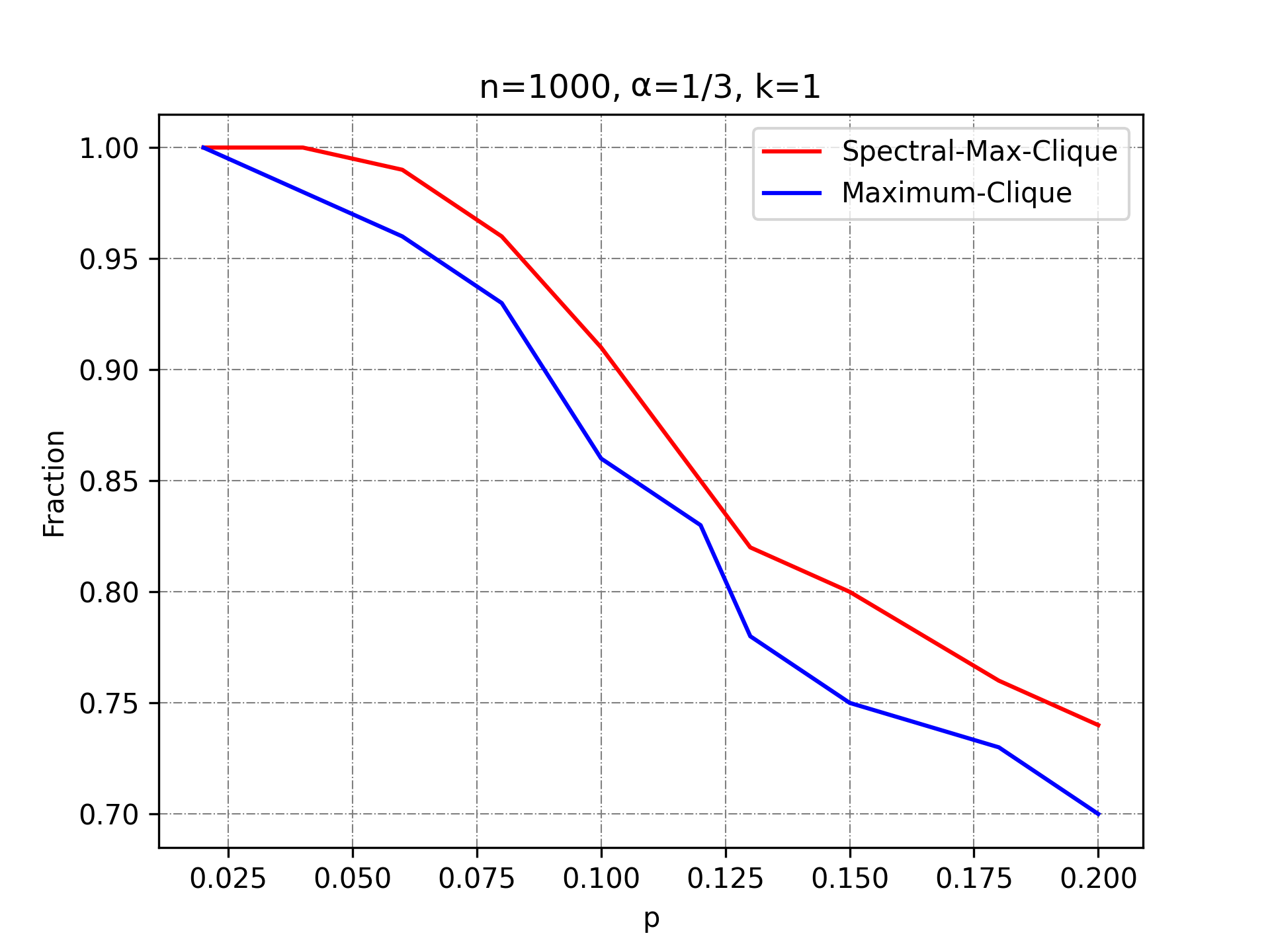}
        \caption{$k = 1$}
    \end{subfigure}
    \quad
    \begin{subfigure}[b]{0.482\textwidth}  
        \centering 
        \includegraphics[width=\textwidth]{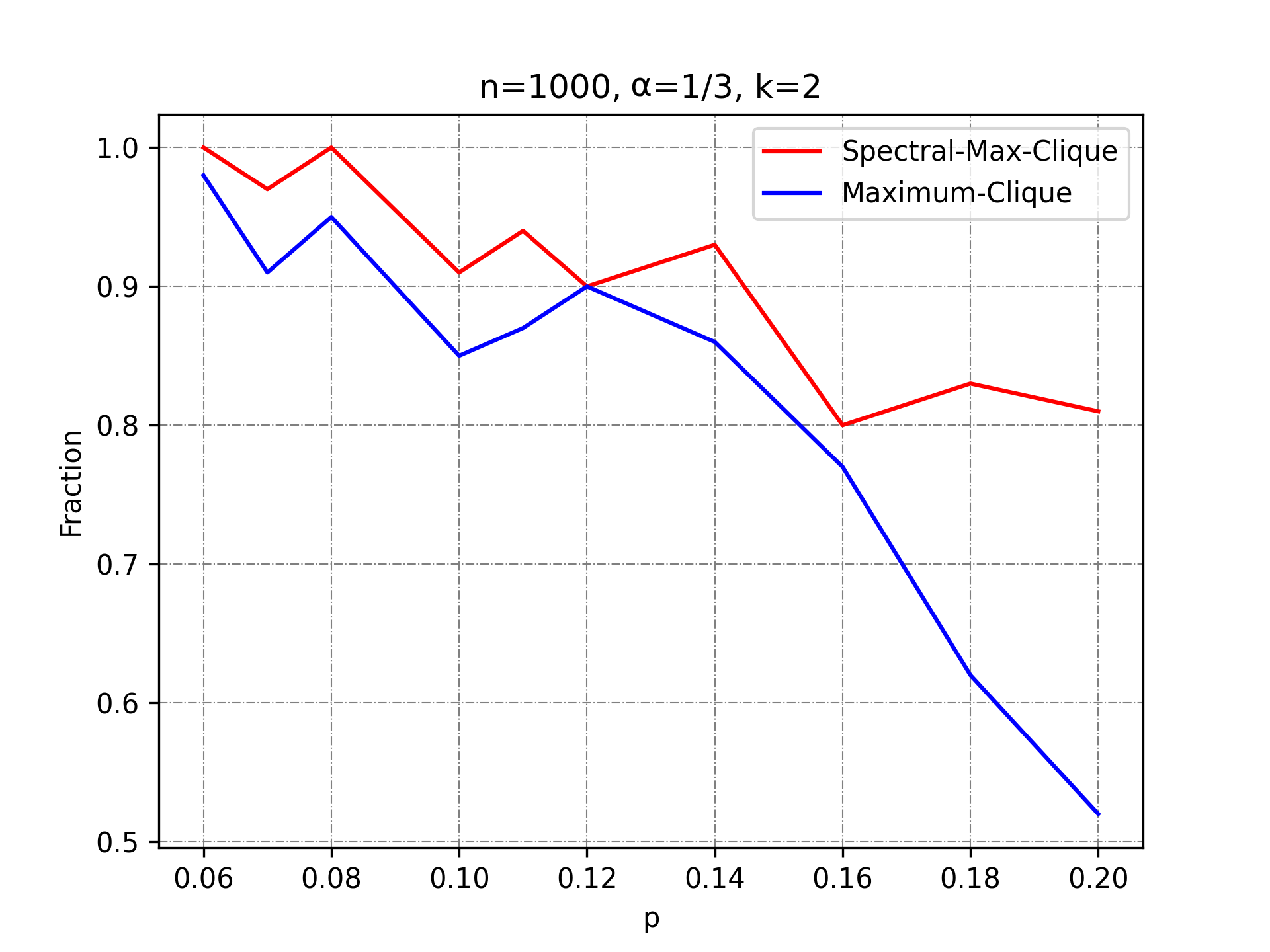}
        \caption{$k = 2$}
    \end{subfigure}
    \vskip\baselineskip
    \begin{subfigure}[b]{0.482\textwidth}   
        \centering 
        \includegraphics[width=\textwidth]{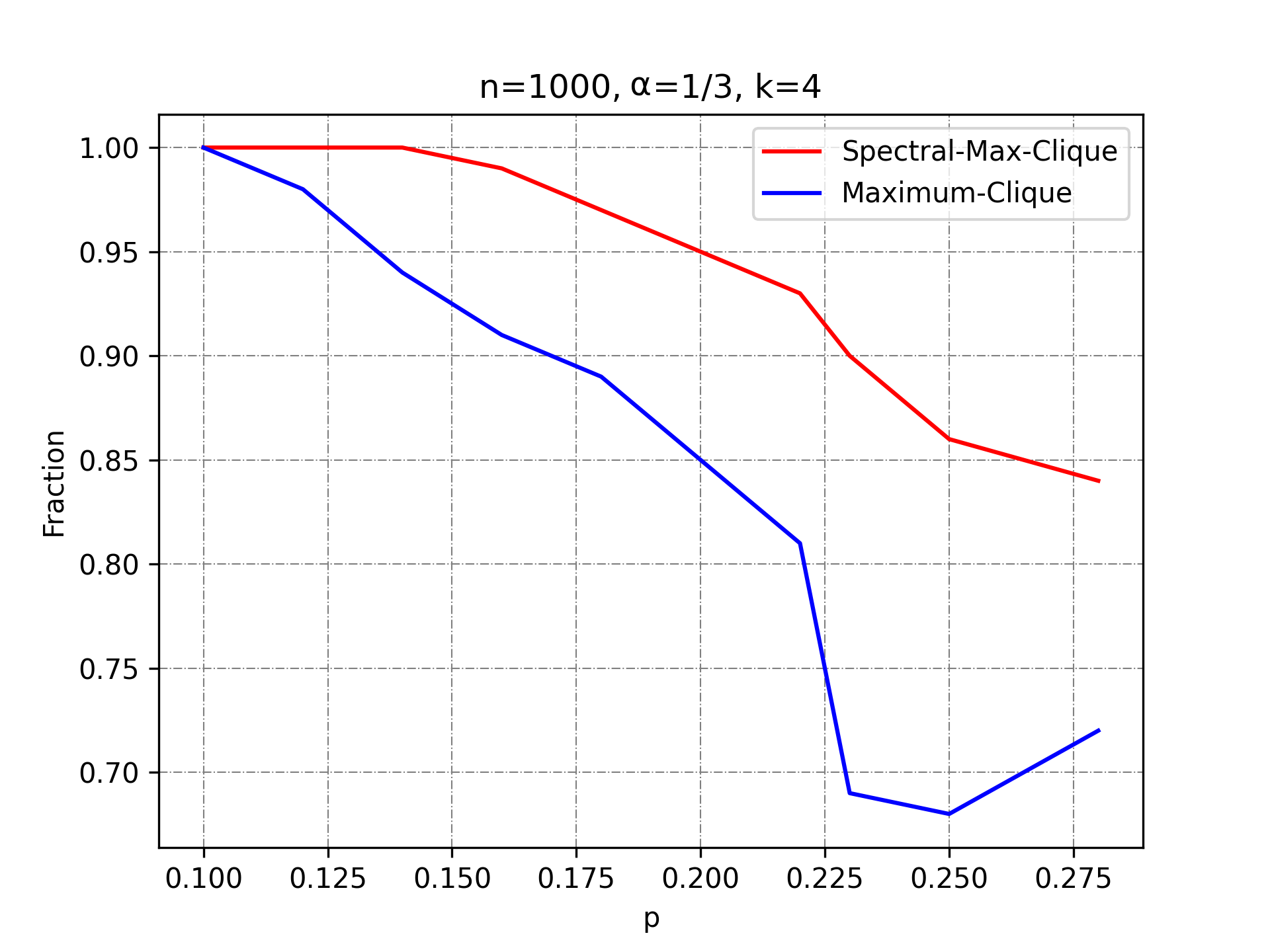}
        \caption{$k = 4$}
    \end{subfigure}
    \quad
    \begin{subfigure}[b]{0.482\textwidth}
        \centering
        \includegraphics[width=\textwidth]{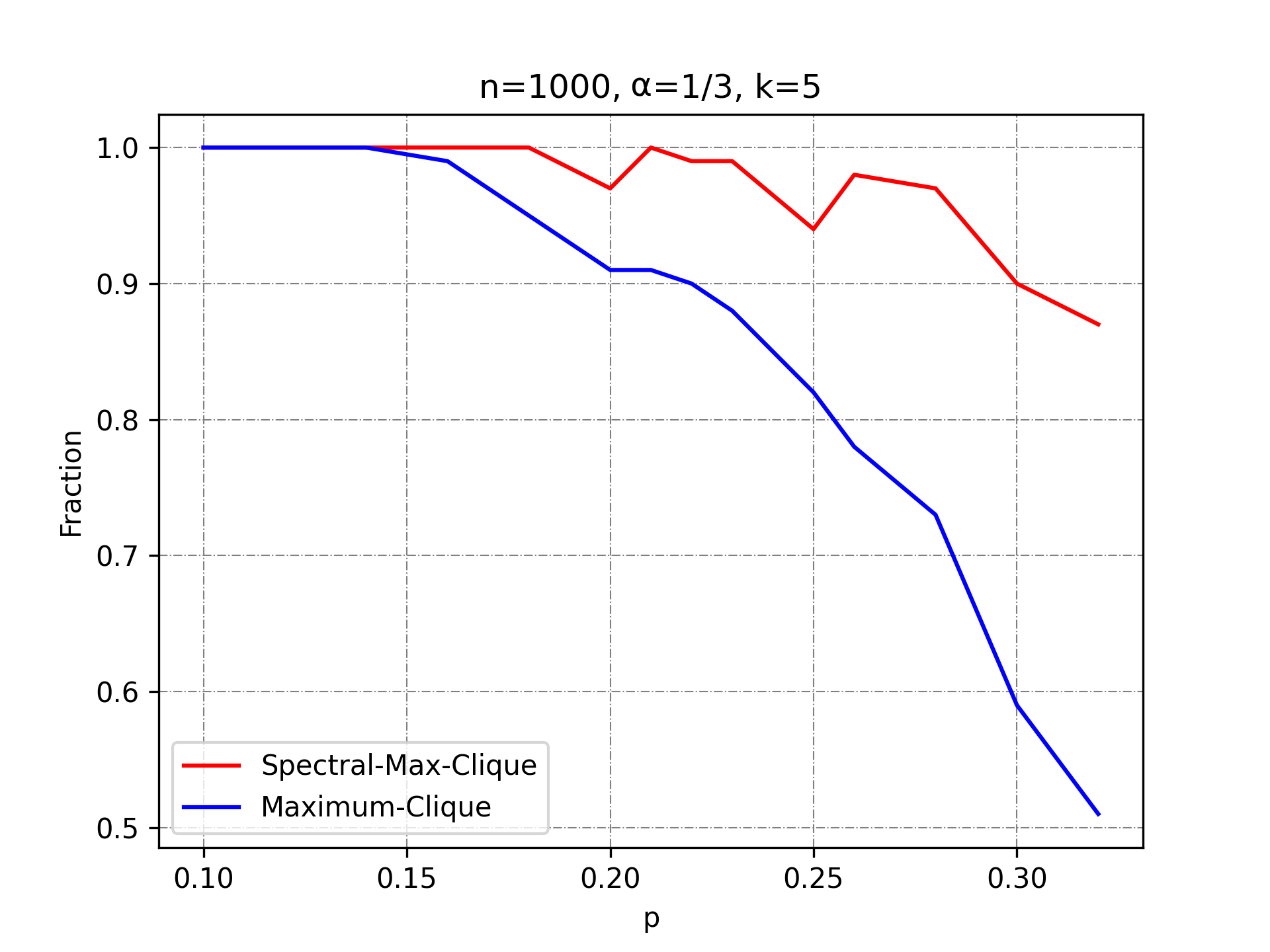}
        \caption{$k = 5$}
    \end{subfigure}
    \quad
    \caption{Approximation guarantee curves for $\alpha = 1/3$, $n = 1000$ and $k=1, 2, 4, 5$.}\label{frac4}
\end{figure*}

\begin{figure*}[ht!]
    \centering
    \begin{subfigure}[b]{0.482\textwidth}
        \centering
        \includegraphics[width=\textwidth]{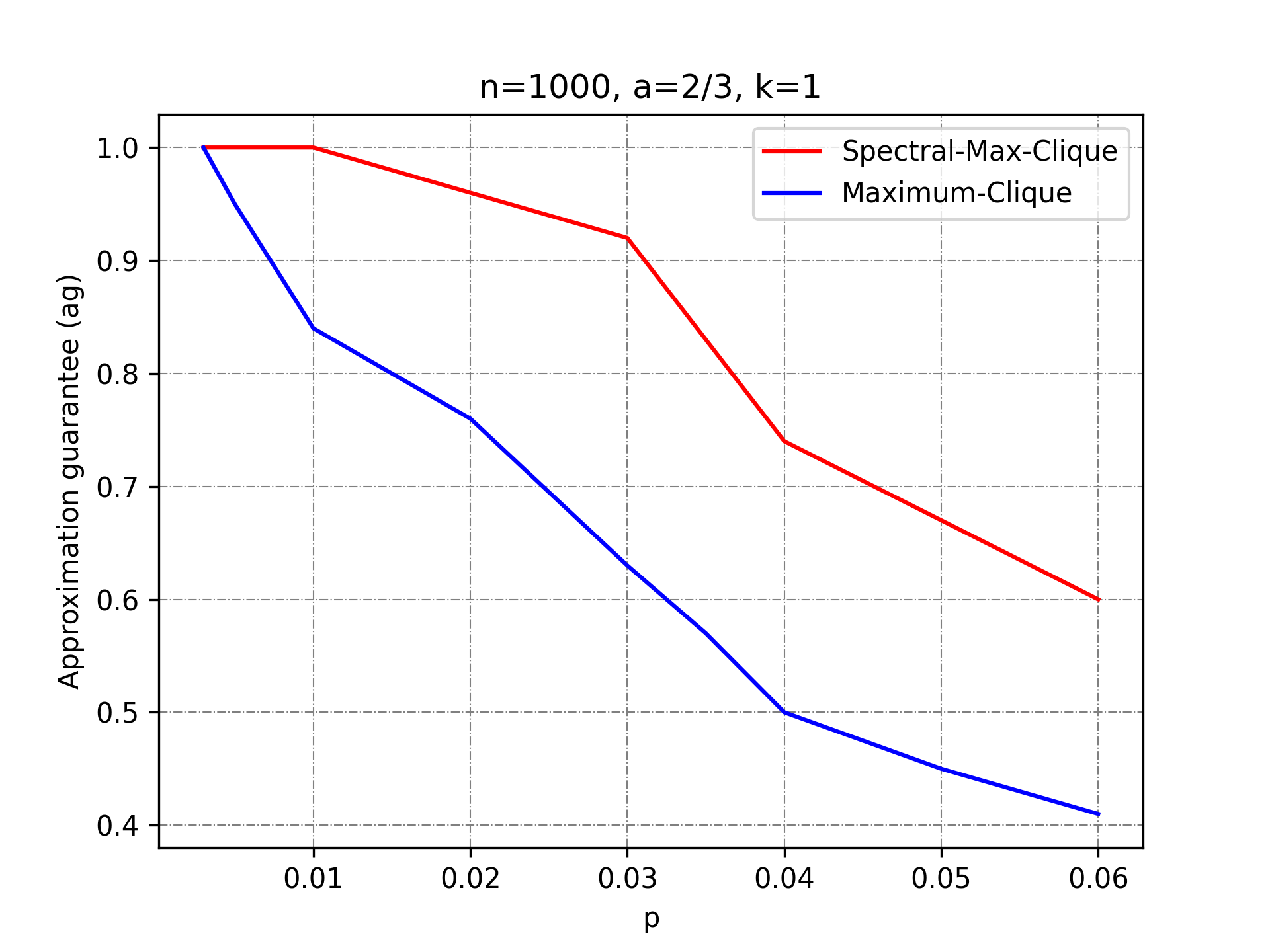}
        \caption{$k = 1$}
    \end{subfigure}
    \quad
    \begin{subfigure}[b]{0.482\textwidth}  
        \centering 
        \includegraphics[width=\textwidth]{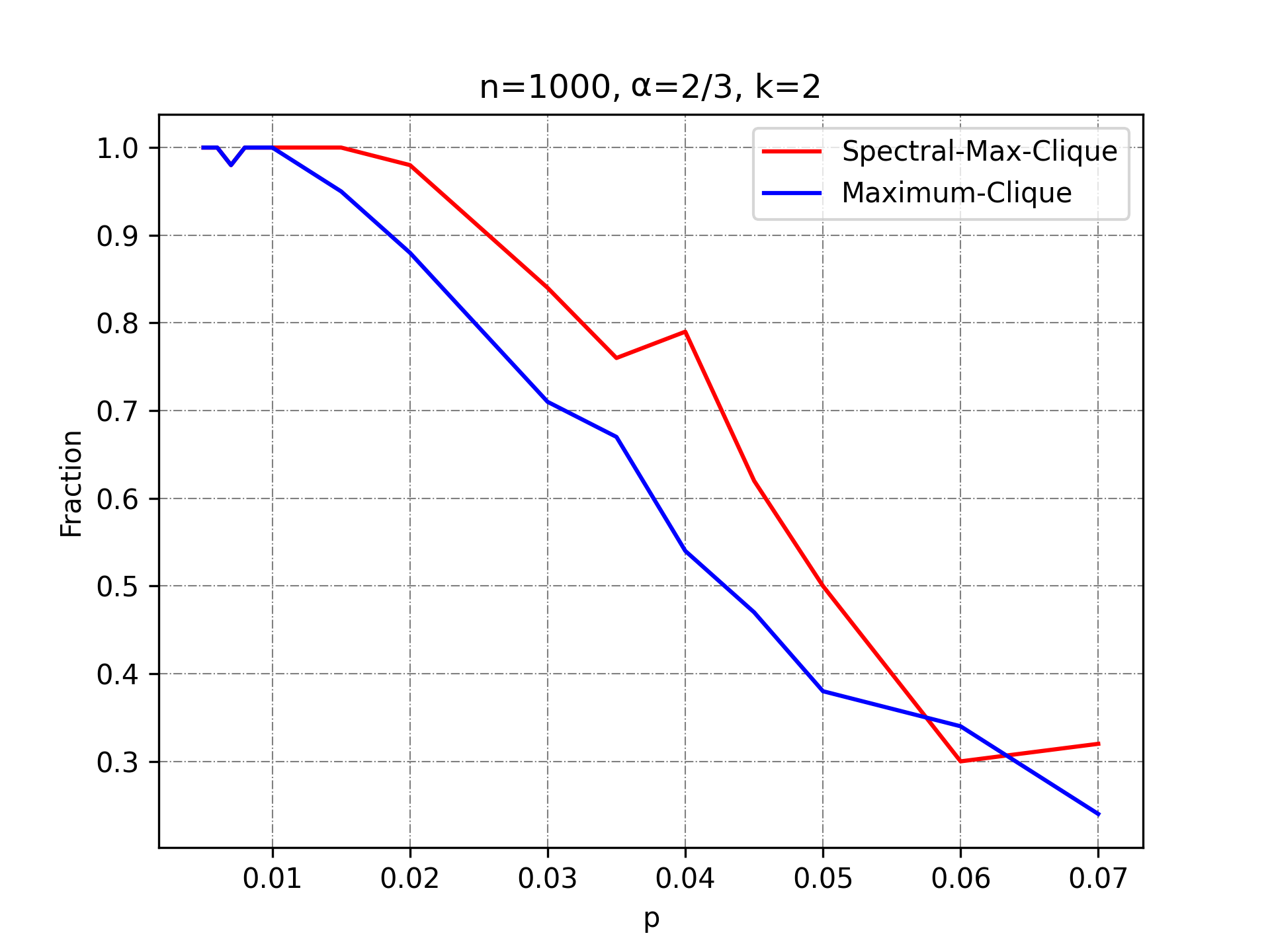}
        \caption{$k = 2$}
    \end{subfigure}
    
    \vskip\baselineskip
    \begin{subfigure}[b]{0.482\textwidth}   
        \centering 
        \includegraphics[width=\textwidth]{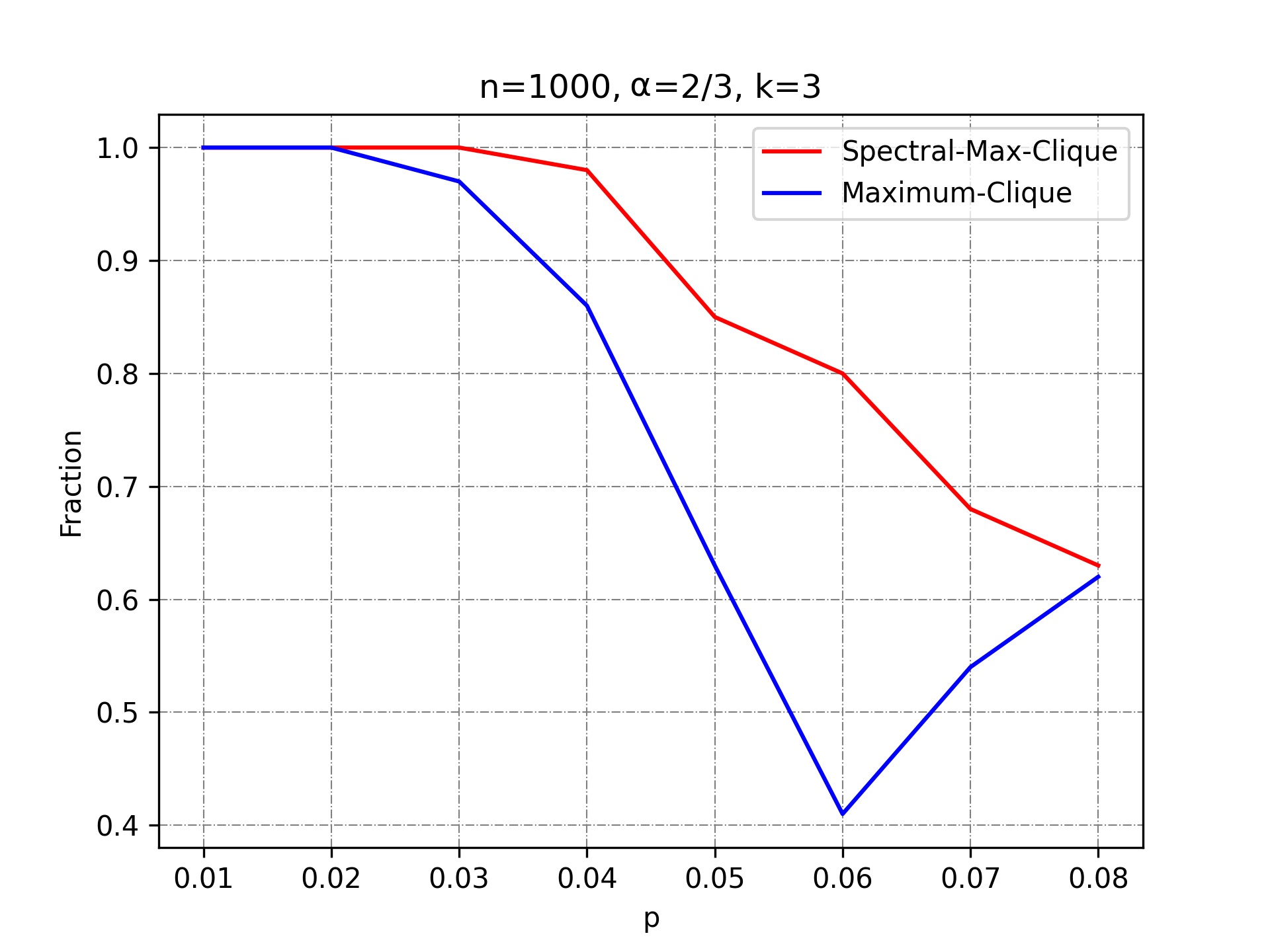}
        \caption{$k = 3$}
    \end{subfigure}
    \quad
    \begin{subfigure}[b]{0.482\textwidth}   
        \centering 
        \includegraphics[width=\textwidth]{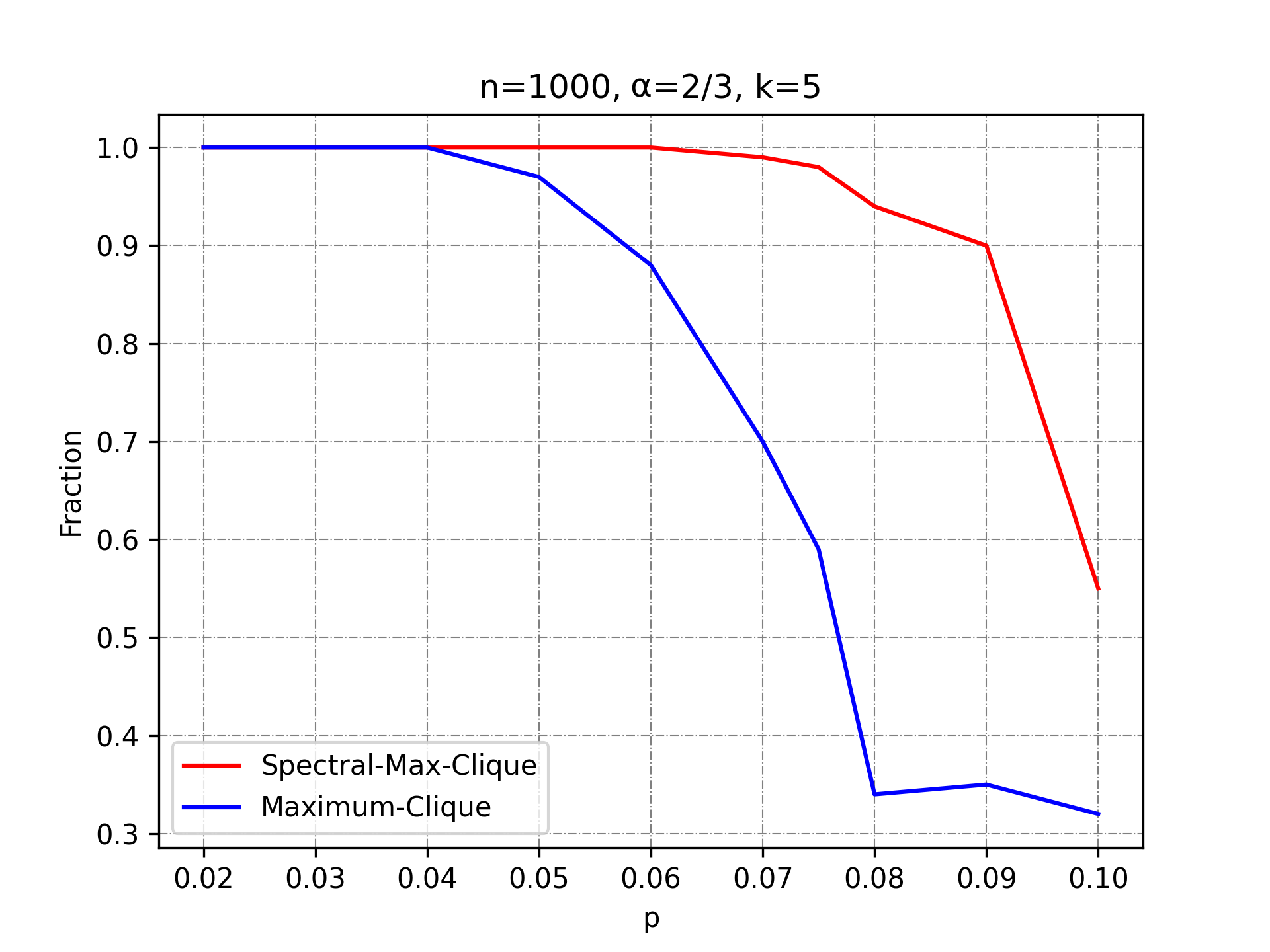}
        \caption{$k = 5$}
    \end{subfigure}
    \vskip\baselineskip
    \begin{subfigure}[b]{0.482\textwidth}   
        \centering 
        \includegraphics[width=\textwidth]{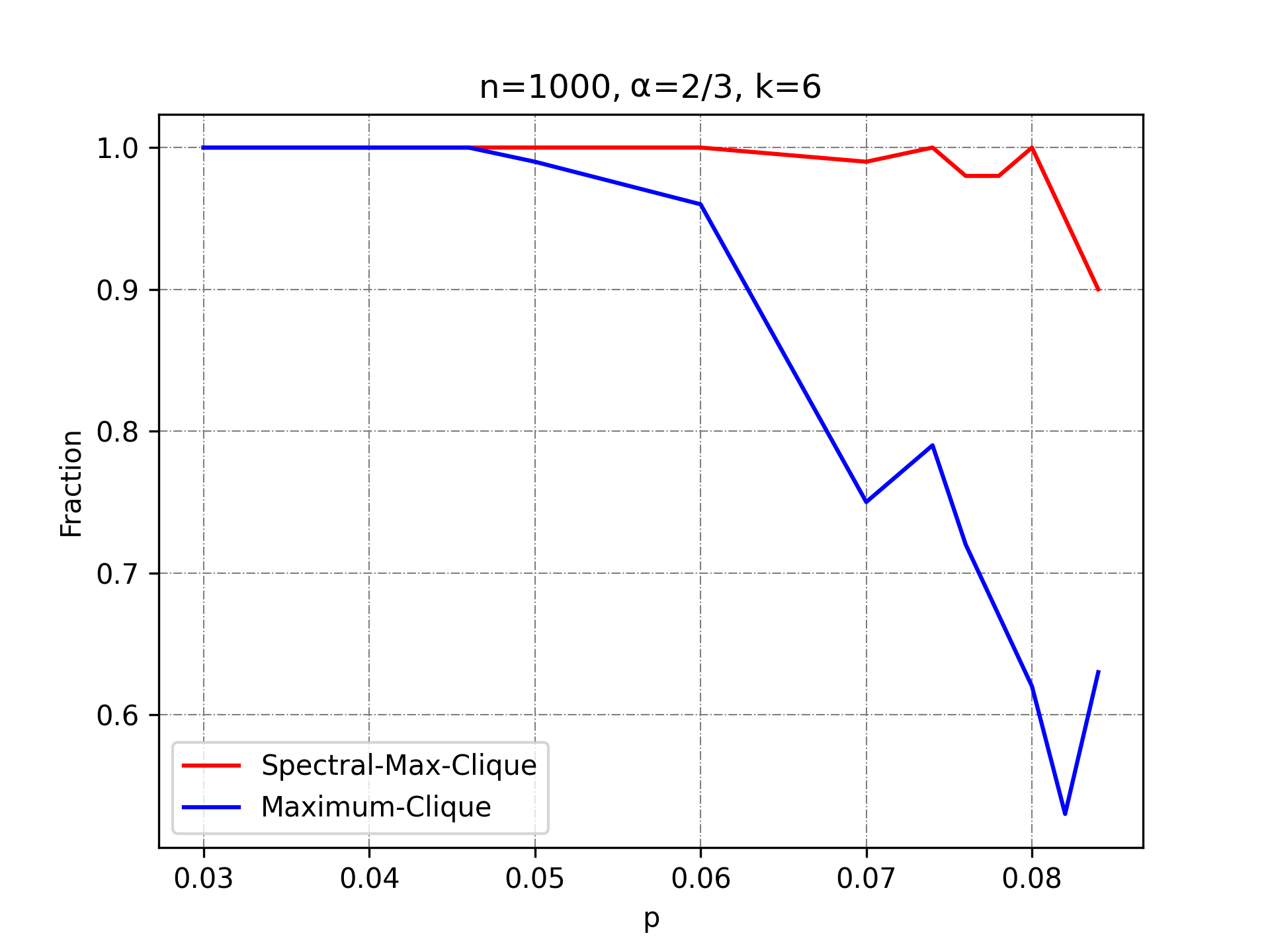}
        \caption{$k = 6$}
    \end{subfigure}
    
    \caption{Approximation guarantee curves for $\alpha = 2/3$, $n = 1000$ and $k=1, 2, 3, 5, 6$.}\label{frac5}
\end{figure*}

\begin{figure*}[ht!]
    \centering
    \begin{subfigure}[b]{0.482\textwidth}
        \centering
        \includegraphics[width=\textwidth]{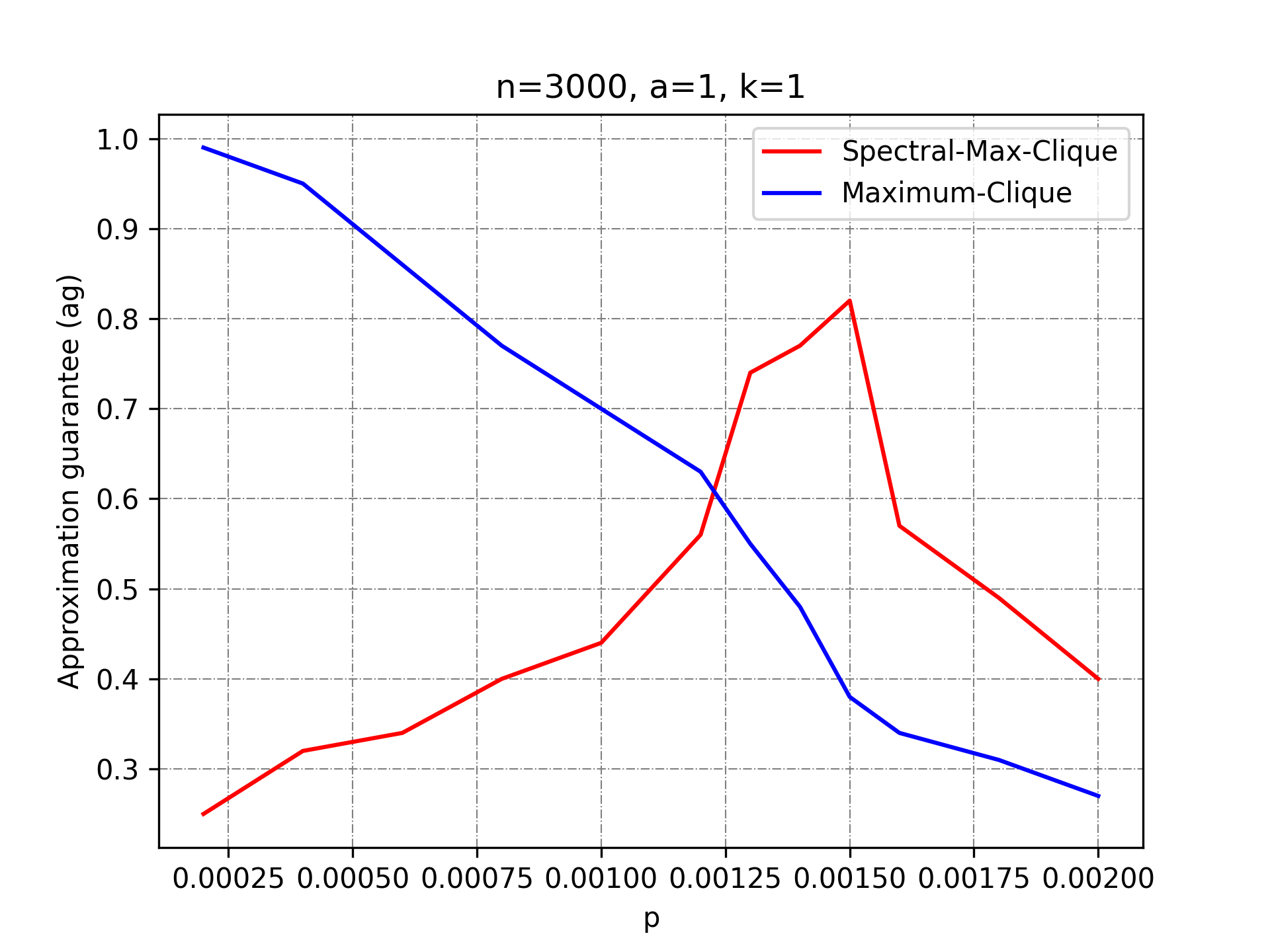}
        \caption{$k = 1$}
    \end{subfigure}
    \quad
    \begin{subfigure}[b]{0.482\textwidth}  
        \centering 
        \includegraphics[width=\textwidth]{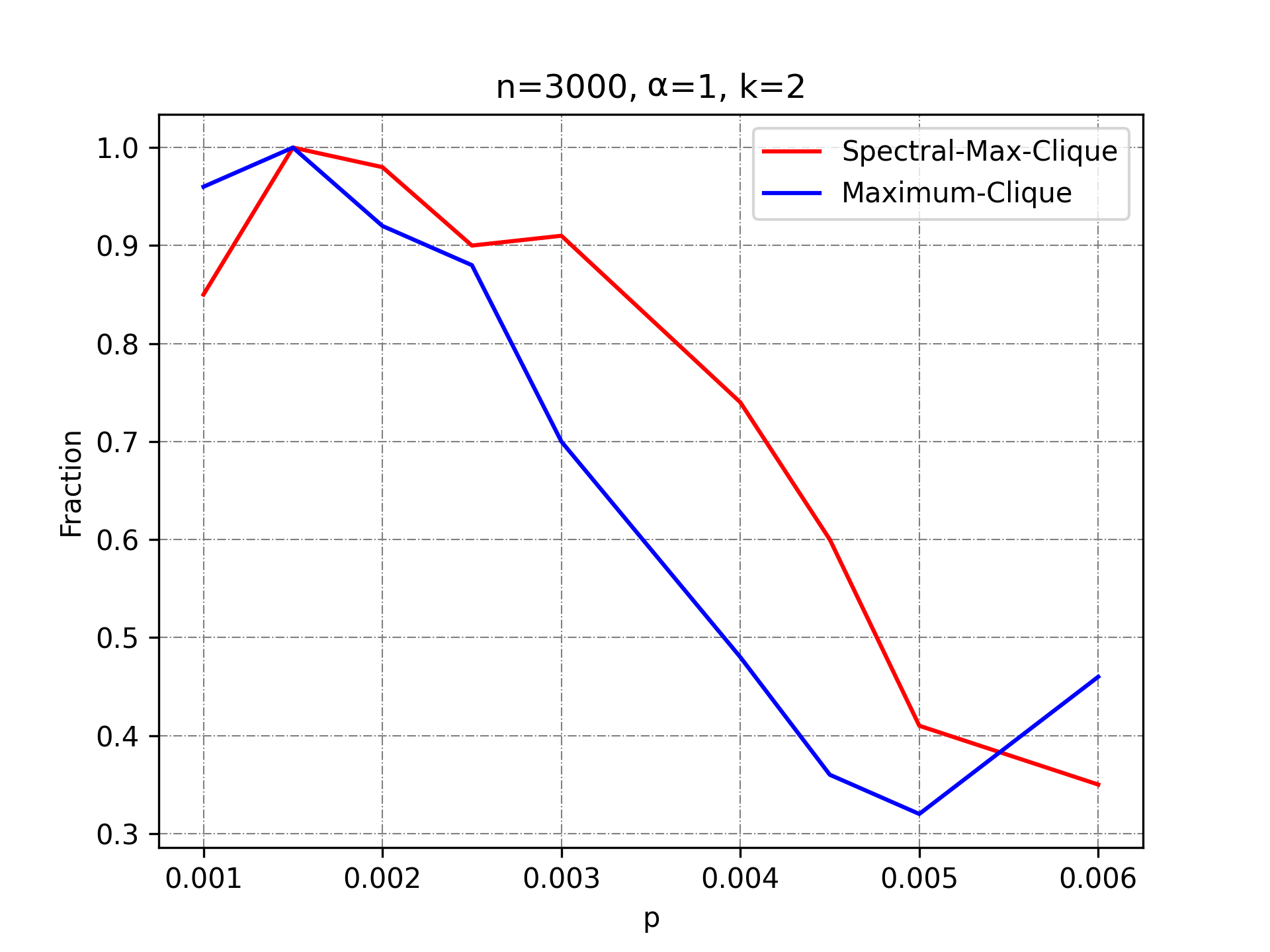}
        \caption{$k = 2$}
    \end{subfigure}
    
    \vskip\baselineskip
    \begin{subfigure}[b]{0.482\textwidth}   
        \centering 
        \includegraphics[width=\textwidth]{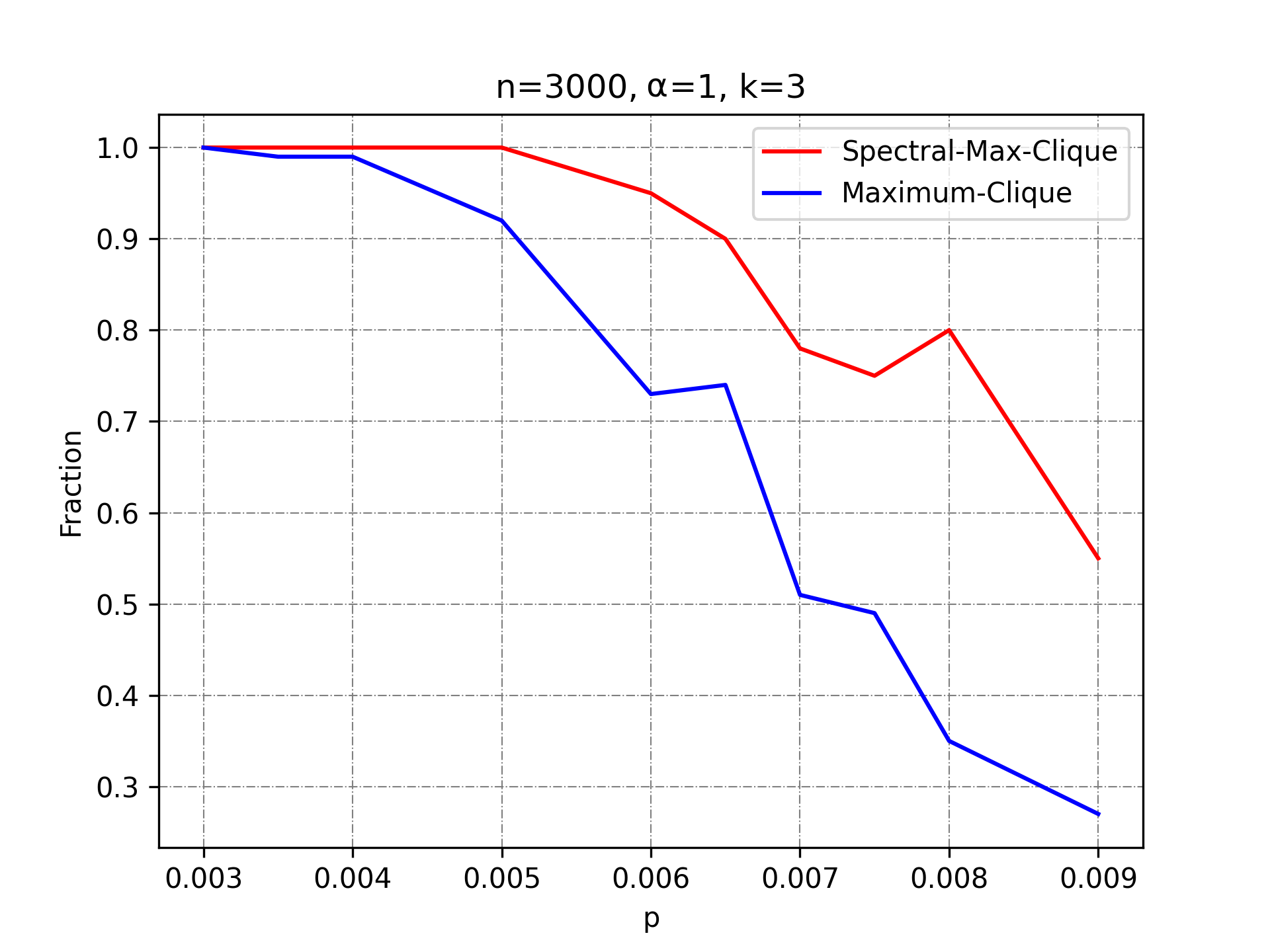}
        \caption{$k = 3$}
    \end{subfigure}
    \quad
    \begin{subfigure}[b]{0.482\textwidth}   
        \centering 
        \includegraphics[width=\textwidth]{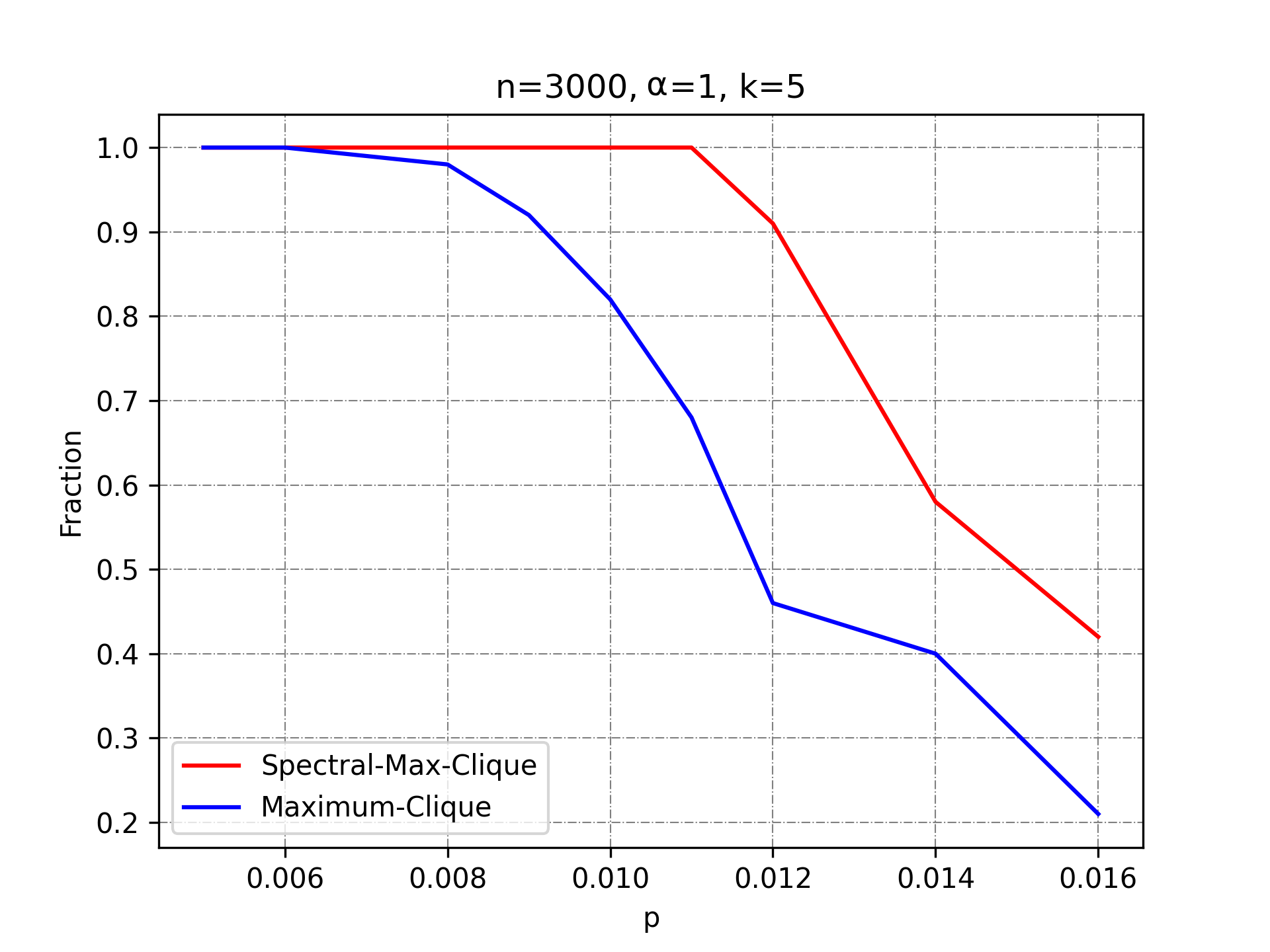}
        \caption{$k = 5$}
    \end{subfigure}
    
    \vskip\baselineskip
    \begin{subfigure}[b]{0.482\textwidth}   
        \centering 
        \includegraphics[width=\textwidth]{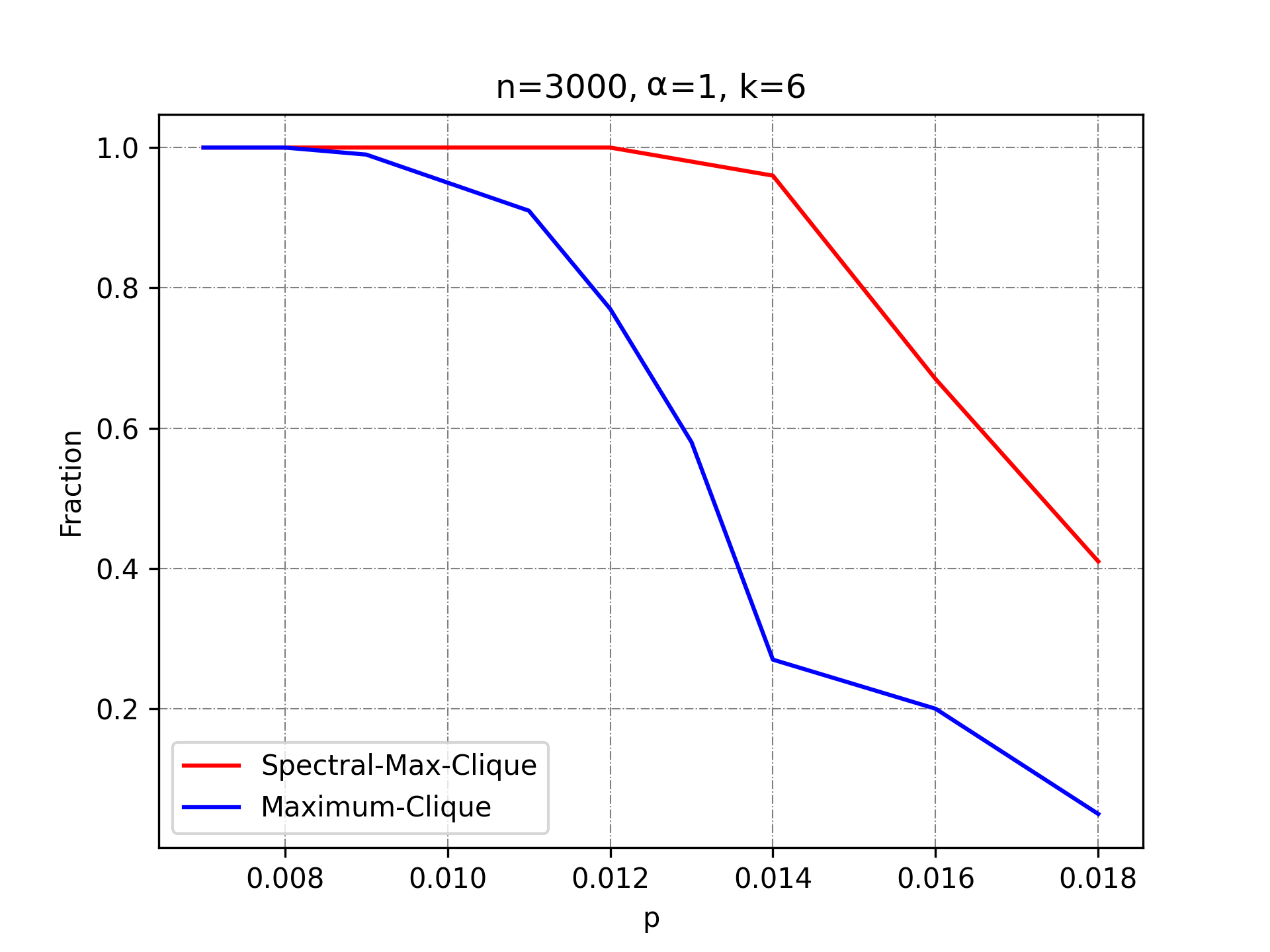}
        \caption{$k = 6$}
    \end{subfigure}
    \quad
    \begin{subfigure}[b]{0.482\textwidth}   
        \centering 
        \includegraphics[width=\textwidth]{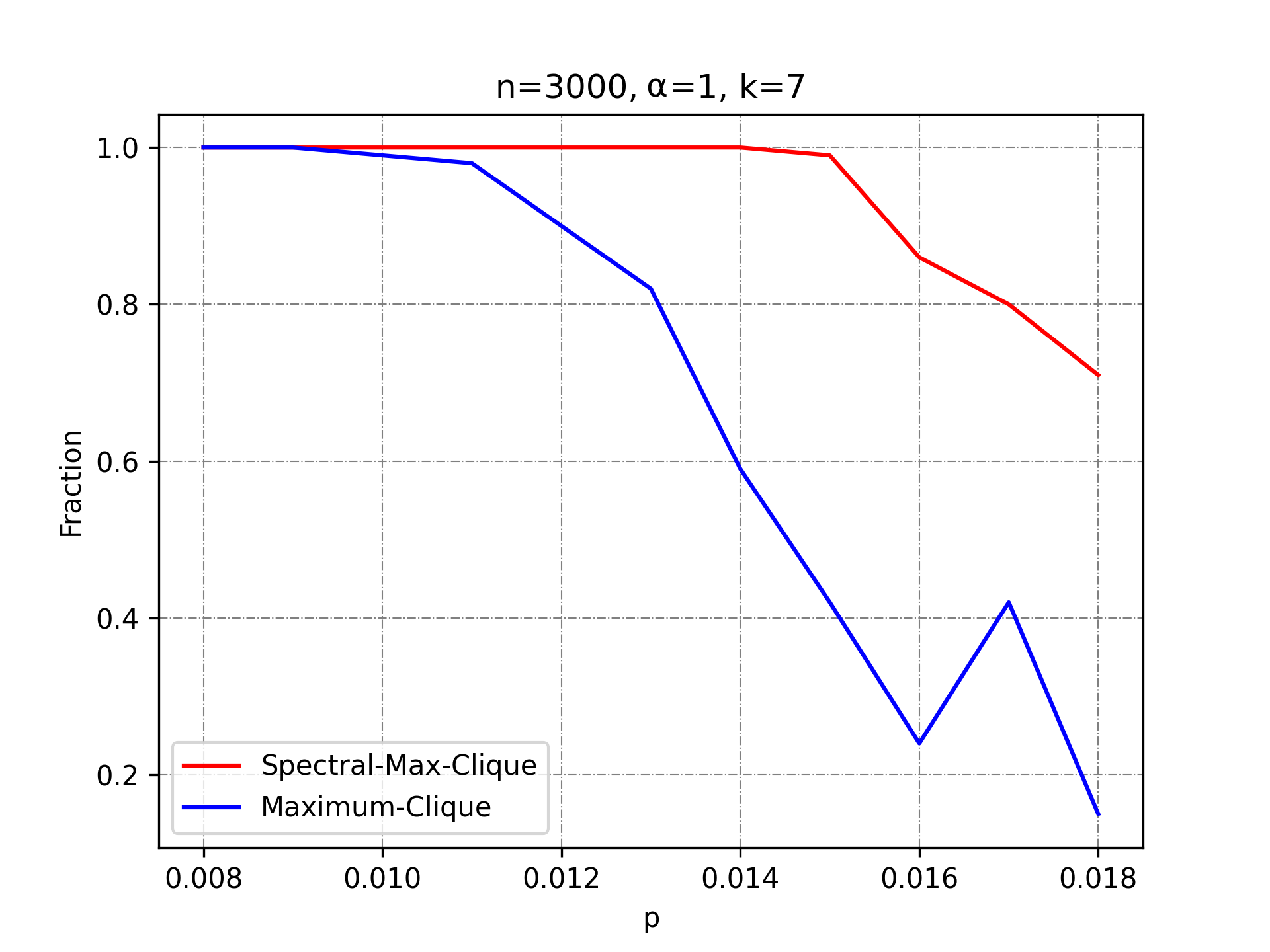}
        \caption{$k = 7$}
    \end{subfigure}
    
    \caption{Approximation guarantee curves for $\alpha = 1$, $n = 3000$ and $k=1, 2, 3, 5, 6, 7$.}\label{frac6}
    
\end{figure*}

In the \cref{fig4,fig5,fig6} we show more of our experimental results, regarding the failure probability curve, while \cref{frac4,frac5,frac6} show the approximation guarantee curve, for different values of $k$. 

%FRACTION FIGURES

\end{document}